\pdfoutput=1

\documentclass[11pt, letterpaper]{article}

\usepackage[nodisplayskipstretch]{setspace}
\usepackage[margin={2cm,2cm}]{geometry} 

\usepackage{sectsty}
\usepackage{abstract}
\usepackage{color}
\usepackage{graphicx}
\usepackage{multicol}
\usepackage{amsmath,amssymb,amstext,amsfonts}
\usepackage[usenames,svgnames]{xcolor}
\usepackage{float}
\usepackage{subfig}
\usepackage{caption}
\usepackage{framed}
\usepackage{authblk}
\usepackage{titlesec}
\usepackage{setspace}
\usepackage{newtxtext,newtxmath,fancyhdr}  
\usepackage[normalem]{ulem} 
\usepackage{enumerate} 
\usepackage{mathrsfs} 
\usepackage{bigints}
\usepackage{bm}
\usepackage{enumitem} 
\usepackage[hypertexnames=false]{hyperref}
\usepackage{wasysym}

\titlespacing*{\section}{0pt}{3ex plus 2ex}{1ex} 
\sectionfont{\fontsize{11}{11}\selectfont} 
\subsectionfont{\fontsize{10}{10}\selectfont} 

\hypersetup{
    colorlinks=true,
    urlcolor=SteelBlue,
    linkcolor=red,
    citecolor=blue,
}

\newcommand*{\Scale}[2][4]{\scalebox{#1}{$#2$}} 

\newcommand{\tinyrmsub}[1]{\mbox{{\tiny{#1}}}}

\newcommand{\rhoh}{\rho_{\mbox{\tiny{H}}}}

\newcommand{\rchi}{\raisebox{\depth}{$\chi$}} 
\newcommand{\PRLsep}{\noindent\makebox[\linewidth]{\resizebox{0.750\linewidth}{1pt}{$\blacklozenge$}}\bigskip}

\begin{document}
\sloppy 
\pagestyle{fancy}
\fancyhead{} 
\fancyhead[OR]{\thepage}
\fancyhead[OC]{{\small{
   \textsf{Vacuum solutions and horizons in ${f(T)}$ gravity}}}}
\fancyfoot{} 
\renewcommand\headrulewidth{0.5pt}
\addtolength{\headheight}{2pt} 
\global\long\def\tdud#1#2#3#4#5{#1_{#2}{}^{#3}{}_{#4}{}^{#5}}
\global\long\def\tudu#1#2#3#4#5{#1^{#2}{}_{#3}{}^{#4}{}_{#5}}

\twocolumn

\title{\vspace{-2cm}\hspace{-0.0cm}\rule{\linewidth}{0.2mm}\\
\bf{\Large{\textsf{On spherically symmetric vacuum solutions and horizons in covariant $\bm{f(T)}$ gravity theory}}}}


\author[1,2]{\small{Andrew DeBenedictis}\thanks{\href{mailto:adebened@sfu.ca}{adebened@sfu.ca}}}%

\author[3]{\small{Sa{\v s}a Iliji{\'c}}\thanks{
   \href{mailto:sasa.ilijic@fer.hr}{sasa.ilijic@fer.hr}}}%

\author[3]{\small{Marko Sossich}\thanks{
   \href{mailto:marko.sossich@fer.hr}{marko.sossich@fer.hr}}}%


\affil[1]{\footnotesize{\it{Simon Fraser University}}\\
   \footnotesize{\it{8888 University Drive, Burnaby, BC, V5A 1S6, Canada}} \protect\\
   \footnotesize{and}
}

\affil[2]{\footnotesize{\it{The Pacific Institute
   for the Mathematical Sciences \vspace{0.2cm}}}
}

\affil[3]{
   \footnotesize{\it{Department of Applied Physics,
   Faculty of Electrical Engineering and Computing, University of Zagreb}}\\
   \footnotesize{\it{HR-10000 Zagreb, Unska 3, Croatia}}
}

\date{\vspace{-0.8cm}({\footnotesize{March 15, 2022}})} 
\twocolumn[ 
  \begin{@twocolumnfalse}  
  \begin{changemargin}{1.75cm}{1.75cm} 
\maketitle
\end{changemargin}
\vspace{-1.0cm}
\begin{changemargin}{1.5cm}{1.5cm} 
\begin{abstract}
{\noindent\small{In this paper we study properties that the vacuum must possess in the minimal extension to the teleparallel equivalent of general relativity (TEGR) where the action is supplemented with a quadratic torsion term. No assumption is made about the weakness of the quadratic term although in the weak-field regime the validity of our previously derived perturbative solution is confirmed. Regarding the exact nature of the vacuum, it is found that if the center of symmetry is to be regular, the mathematical conditions on the tetrad at the isotropy point mimic those of general relativity. With respect to horizons it is found that, under very mild assumptions, a smooth horizon cannot exist unless the quadratic torsion coupling, $\alpha$, vanishes, which is the TEGR limit (with the Schwarzschild tetrad as its solution). This analysis is then supplemented with computational work utilizing asymptotically Schwarzschild boundary data. It is verified that in no case studied does a smooth horizon form. For $\alpha > 0$ naked singularities occur which break down the equations of motion before a horizon can form. For $\alpha < 0$ there is a limited range of $\alpha$ where a vacuum horizon might exist but, if present, the horizon is singular. Therefore physically acceptable black hole horizons are problematic in the studied theory at least within the realm of vacuum static spherical symmetry. These results also imply that static spherical matter distributions generally must have extra restrictions on their spatial extent and stress-energy bounds so as to render the vacuum solution invalid in the singular region and make the solutions finite.}}
\end{abstract}
\noindent{\footnotesize PACS(2010): 02.40.Xx \;\, 04.50.Kd}
\rule{\linewidth}{0.2mm}
\end{changemargin}
\end{@twocolumnfalse} 
]
\saythanks 
\vspace{0.5cm}
{\setstretch{0.9} 
\section{Introduction}
General relativity is now known to be a highly successful theory of gravity on many energy scales. It has passed a number of solar system tests \cite{ref:grtests} and, more recently, it has been shown to be accurate even in high energy scenarios \cite{ref:bhmerger}. General relativity is based on the concept of intrinsic curvature as the cause of the gravitational field. There also exists a much less known, but completely equivalent theory of gravity based on torsion and no curvature. This theory is known as the teleparallel equivalent of general relativity (TEGR). Instead of an action constructed linear in the Ricci scalar with the metric as the degree of freedom, TEGR instead is derived from an action constructed linearly from the torsion scalar with the tetrad being the degree of freedom. The two theories yield exactly the same equations of motion, save for a difference in a boundary term (and hence a possible difference in junction conditions \cite{ref:Dunsby}, \cite{ref:jesse}). Since the two theories are equivalent it is a matter of choice which theory one chooses to work with, provided the choice is between general relativity and TEGR.

However, it is possible that the full theory is not general relativity or TEGR, but only yields these in a certain limit. In curvature based theories arguably the most popular extension to the Einstein-Hilbert Lagrangian density is one where the Ricci scalar in the action is supplemented with a term quadratic in the Ricci scalar. The quadratic term is sometimes referred to as the Starobinsky term \cite{ref:staro}. This ``$R+\alpha R^{2}$'' theory could be viewed as the correct full theory of curvature gravity, or just the first two terms in a Lagrangian density which is power-expandable in powers of the curvature scalar about small curvature. In the latter case the $R^{2}$ term is seen as a correction due to a more general $f(R)$ gravity theory.

One can do the same type of extension in the torsion theory, giving rise to what is known as $f(T)$ gravity theory, or extended teleparallel gravity. The resulting $f(T)$ equations of motion will no longer mimic those of the corresponding curvature theory beyond linear order in the action and so the two theories will generally make different predictions. One feature of $f(T)$ gravity is that the differential equations of motion retain their second-order nature even when $f(T)$ is no longer simply linear in the torsion scalar, whereas $f(R)$ becomes a fourth-order theory beyond the linear Lagrangian density. 

The covariant theory (in this manuscript meaning  with spin connection explicitly included) produces the same equations of motion as the pure tetrad theory if one chooses a ``good'' tetrad  \cite{ref:tamgood} in the pure tetrad theory. Therefore, using a good tetrad with no spin connection yields the same equations of motion as the ones here. A good tetrad in the pure tetrad frame does yield locally covariant equations of motion provided that, when one locally Lorentz transforms the tetrad to another frame, one must also pick up the proper non-zero spin connection in this new frame. 

The $f(T)$ gravity is not nearly as well studied as its $f(R)$ counterpart but interest has increased dramatically in the past couple of decades when considering candidates for modified gravity theories. The greatest amount of work in teleparallel gravity has arguably been performed in the arena of cosmology \cite{ref:sharifandrani} - \cite{ref:dambrosio}. There it has been shown that such modifications to the gravitational action  may be able to naturally produce dark matter and dark energy effects \cite{ref:dmstart}-\cite{ref:dmend}. Stellar structure has also been studied in some detail \cite{ref:tamgood} - \cite{ref:fortes} as well as black holes \cite{ref:bh1} - \cite{ref:nashsari}. A nice review of the subject may be found in \cite{ref:jacksonlevi}.

\section[A brief review of covariant $f(T)$ gravity]{A brief review of covariant $\bm{f(T)}$ gravity}

In this manuscript we will refer to the Riemann tensor specifically constructed from the Levi-Civita connection as the Riemann-Christoffel tensor (although we should caution that sometimes in the mathematical literature this nomenclature refers to the Riemann tensor for any connection). The Riemann-Weitzenb\"{o}ck tensor, whose components are all identically zero, refers to the Riemann tensor constructed specifically from the Weitzenb\"{o}ck connection.

The action for $f(T)$ gravity theory is given by \footnote{Indices are such that hatted Greek letters represent orthonormal indices whereas unadorned Greek letters represent spacetime coordinate indices.} 
\begin{equation}
S=\int \left(\frac{1}{16\pi}\, f(T)  +\mathcal{L}\tinyrmsub{matter}\right){\mbox{det}(h^{\hat{\alpha}}_{\;\;\mu})}\,d^{4}x\,. \label{eq:tegraction}
\end{equation}
Here $h^{\hat{\alpha}}_{\;\;\mu}$ represents the tetrad, which satisfies the condition of metric compatibility:
\begin{equation}
h^{\hat{\alpha}}_{\;\;\mu}h_{\hat{\alpha}\nu}=g_{\mu\nu}\,, \label{eq:metcomp}
\end{equation}
and $f(T)$ is some function of the torsion scalar, $T$, which is constructed out of the torsion tensor, $T^{\alpha}_{\;\;\beta\gamma}$. The torsion is defined from the commutator of the Weitzenb\"{o}ck connection $\Gamma^{\sigma}_{\;\;\beta\gamma}$ with the spin connection, $\omega^{\hat{\alpha}}_{\;\;\hat{\beta}\sigma}$ as
{\allowdisplaybreaks\begin{align}
T^{\hat{\alpha}}_{\;\;\mu\nu}=&h^{\hat{\alpha}}_{\;\;\sigma}\left(\Gamma^{\sigma}_{\;\;\nu\mu}-\Gamma^{\sigma}_{\;\;\mu \nu}\right) := \partial_{\mu}h^{\hat{\alpha}}_{\;\;\nu} - \partial_{\nu}h^{\hat{\alpha}}_{\;\;\mu} \nonumber \\
& + \omega^{\hat{\alpha}}_{\;\;\hat{\beta}\mu}h^{\hat{\beta}}_{\;\;\nu} - \omega^{\hat{\alpha}}_{\;\;\hat{\beta}\nu}h^{\hat{\beta}}_{\;\;\mu}\,.\label{eq:gentor}
\end{align}}
The torsion scalar itself is formed via:
\begin{equation} 
    T :=
    \frac14  T_{\alpha\beta\gamma}  T^{\alpha\beta\gamma}
     + \frac12  T_{\alpha\beta\gamma}  T^{\gamma\beta\alpha}
     - T_{\alpha\beta}^{\;\;\;\alpha}T^{\gamma\beta}_{\;\;\;\;\gamma}\,, \label{eq:genT}
\end{equation}

Even though it is not a tensor we \emph{define} the raising and lowering of indices on the spin connection in the usual way
\begin{equation}
\omega^{\hat{\alpha}}_{\;\;\hat{\beta}\mu}:=g^{\hat{\alpha}\hat{\gamma}}\omega_{\hat{\gamma}\hat{\beta}\mu}\,,\; \omega_{\hat{\gamma}\hat{\beta}}^{\;\;\;\;\nu}:= 
\omega_{\hat{\gamma}\hat{\beta}\mu} g^{\mu\nu},\, \mbox{etc.}\,, \nonumber
\end{equation}
the hatted metric being the orthonormal metric.

The equations of motion result from extremizing the action (\ref{eq:tegraction}) with respect to the tetrad $h^{\hat{\alpha}}_{\;\;\mu}$  yielding
\begin{align}
&\frac12 \, g_{\mu\nu} f(T)
+ \frac{\mathrm{d} f(T)}{\mathrm{d} T}
   \left( \mathring{G}_{\mu\nu} - \frac{1}{2} \, g_{\mu\nu} T \right) \nonumber \\
&+ \frac{\mathrm{d}^2 f(T)}{\mathrm{d} T^2} \,
	S_{\mu\nu}{}^{\lambda} \, \partial_{\lambda} T
= 8\pi \, \mathcal{T}_{\mu\nu}, \label{eq:eoms}
\end{align}
where $\mathcal{T}_{\mu\nu}$ represents the components of the symmetric stress-energy tensor, which will be set to zero here as we will be dealing with vacuum solutions. $\mathring{G}^{\mu\nu}$ is the Einstein tensor, constructed from the Ricci scalar and Ricci tensor created from the Christoffel connection. We will use a ring over quantities constructed from the Christoffel connection. The quantity $S^{\mu\nu\rho}$ is known as the superpotential, and is given by
\begin{equation}
S_{\alpha\mu\nu} =  K_{\mu\nu\alpha}
               - g_{\alpha\nu} T^{\lambda}{}_{\mu\lambda}
               + g_{\alpha\mu} T^{\lambda}{}_{\nu\lambda} \,, \label{eq:superpot}
\end{equation}
with $K_{\mu\nu\alpha}$ the contorsion
(sometimes referred to as contortion) tensor defined by
\begin{equation}
K_{\alpha\mu\nu} = \frac12 \big(
T_{\nu\alpha\mu} + T_{\mu\alpha\nu} - T_{\alpha\mu\nu} \big)\,. \label{eq:contor}
\end{equation}

The form that we have written equations (\ref{eq:eoms}) is not the common way that they are usually found in the $f(T)$ literature, but they are equivalent \cite{ref:bianchi}. The form in (\ref{eq:eoms}) makes it particularly convenient to compare $f(T)$ gravity with the Einstein equations of general relativity and isolate the differences in the two theories. One can see in (\ref{eq:eoms}) that when taking $f(T)=T$ (TEGR) one recovers readily the Einstein equations.

The primary role of the spin connection is to render the theory locally Lorentz covariant \cite{ref:goodbadtets} - \cite{ref:golov2}.
If the spin connection is ignored it is known that generally the resulting $f(T)$ theory is not covariant under local Lorentz transformations. It is still possible to achieve physically sensible equations of motion without the spin connection but one must then choose a tetrad which yields zero for all components of the spin connection. Such a tetrad, often referred to as a ``good'' tetrad in the $f(T)$ literature \cite{ref:goodbadtets} - \cite{ref:davood},  will then yield physically appropriate equations of motion. These equations of motion will then be identical to the equations of motion created with an arbitrary metric compatible tetrad (meaning (\ref{eq:metcomp}) is satisfied) but without ignoring the spin connection. To reiterate, if one chooses to ignore the spin connection one cannot use just any metric compatible tetrad, but must choose one that yields zero spin connection, whereas with a properly computed spin connection any metric compatible tetrad may be utilized.  It is generally simpler, and safer, to include the spin connection so that one does not need to worry about local Lorentz invariance. The drawback to this is that at this time there is no scheme to calculate the appropriate inertial spin connection for general scenarios. 

Although it is still not clear how to calculate the appropriate inertial spin connection in general cases, there has been some good progress on this in the past few years  \cite{ref:KandS}, \cite{ref:lomonosov}. The methods presented in \cite{ref:KandS} and \cite{ref:lomonosov} provide slightly different prescriptions on how to isolate the inertial (versus the truly gravitational) degrees of freedom, and provide a method to compute the spin connection so that it renders the equations of motion (\ref{eq:eoms}) Lorentz covariant and therefore truly describing gravitational effects only. The schemes do not work in all scenarios. As an example, the method does not yield satisfactory equations of motion when $A$ and $B$ are time also dependent. For the static spherical scenario however the method is robust and they work quite well for the case of static spherical symmetry. 

In the method of \cite{ref:KandS} one can compute the appropriate spin connection by first considering a tetrad ansatz of choice in (\ref{eq:gentor}). For example, relevant to this work, a tetrad compatible with spherical symmetry is chosen. One then takes the $G\rightarrow 0$ (gravitational constant) limit in the resulting torsion tensor, and sets this torsion tensor equal to zero. This results in a set of equations for the spin-connection components which one must solve.

The method in \cite{ref:lomonosov} differs slightly in that one first computes the spin connections via
\begin{equation}
\omega^{\hat{\alpha}}{}_{\hat{\beta}\mu} = - (\mathring{\nabla}_{\mu} h^{\hat{\alpha}}{}_{\nu}) h_{\hat{\beta}}{}^{\nu}
= - ( \partial_{\mu} h^{\hat{\alpha}}{}_{\nu}
      - \mathring{\Gamma}^{\lambda}{}_{\nu\mu} h^{\hat{\alpha}}{}_{\lambda} ) h_{\hat{\beta}}{}^{\nu}
\,,
\end{equation}
using the tetrad ansatz (eg. spherical symmetry) and the Levi-Civita (Christoffel) connection, $\mathring{\Gamma}^{\mu}_{\;\nu\sigma}$. Then the flat space limit (in the Riemann-Christoffel sense) is set in the resulting expression and this yields the components of the inertial spin connection.

Below we will utilize these methods to compute the necessary inertial spin connection coefficients. We should mention here that the spin connection computed according to these methods yields the correct inertial spin connection required for full local Lorentz invariance, which includes parity and time-reversal, that produces zero torsion for Minkowski spacetime. This is demanded, for example, so that spinors do not experience gravitational effects in Minkowski spacetime. It may be possible in $f(T)$ gravity to just demand that spinors do not couple to the torsion, but since in general such coupling naturally arises in spinor theory we do not consider scenarios which yield non-zero torsion in Minkowski spacetime as it is currently not fully known what all the repercussions are of having torsion in Minkowski spacetime. These are stronger conditions than simply having the equations of motion (\ref{eq:eoms}) be symmetric. The resulting equations of motions must be symmetric as well as be locally Lorentz invariant (including the discrete transformations on the tetrad of parity and time reversal) in order to have covariant $f(T)$ gravity, and the methods of \cite{ref:KandS} and \cite{ref:lomonosov} provide the spin connection that achieves both these criteria in static spherical symmetry.


\section[Possible measures of regularity in f(T) gravity]{Possible measures of regularity in $\bm{f(T)}$ gravity}
Regularity in $f(T)$ gravity is rather a trickier issue than in curvature based theories such as general relativity. For example, in curvature theories one possesses a curvature singularity wherever at least one of the orthonormal components of the Riemann curvature tensor becomes infinite. In $f(T)$ gravity though the Riemann-Weitzenb\"{o}ck tensor is identically zero. One may perhaps then appeal to the fact that the tensor analogous to the Riemann curvature tensor in $f(T)$ gravity is the torsion tensor (\ref{eq:gentor}). This tensor however does not provide a reliable diagnostic of physical pathologies in the spacetime. One way to see this is to construct the torsion tensor with the Schwarzschild solution's tetrad. The Schwarzschild solution is the unique spherically symmetric vacuum solution in TEGR, which is a valid theory within the realm of $f(T)$ gravities. Therefore the Schwarzschild horizon is a bona-fide physically acceptable black hole horizon in $f(T)=T$ gravity theory. It can be readily verified though that some components of the torsion tensor, both in the coordinate and orthonormal frames, diverge on the Schwarzschild horizon, even though it is well known that this surface is benign in TEGR and general relativity. The same is true of the superpotential (\ref{eq:superpot}) and contortion tensor (\ref{eq:contor}); again some coordinate and orthonormal components of these tensors diverge on the benign Schwarzschild horizon. The torsion scalar itself also sheds no light on regularity, since it is also infinite on the Schwarzschild horizon. Similarly, the scalars $T_{\alpha\beta\gamma}T^{\alpha\beta\gamma}$, $K_{\alpha\beta\gamma}K^{\alpha\beta\gamma}$, and $S_{\alpha\beta\gamma}S^{\alpha\beta\gamma}$ diverge on the Schwarzschild horizon and so also do not provide a good benchmark for true singular behavior.

In this paper we will make clear specifically what is  meant by ``singularity'' or ``regularity'' in the sections where the issue arises, but to summarize we generally mean that the equations of motion themselves are ill or well behaved in some sense. We also will appeal to the Riemann-Christoffel tensor in the orthonormal frame, or the Riemann-Christoffel Kretschmann scalar, $\mathring{R}_{\alpha\beta\gamma\delta}\mathring{R}^{\alpha\beta\gamma\delta}$, being finite or not. As mentioned above, the Riemann-Christoffel criteria may seem peculiar in a theory whose spacetime connection is not the Christoffel connection. The reason we sometimes utilize this condition as a measure of the spacetime's regularity is that even in $f(T)$ gravity the paths of free-falling particles are governed by the geodesic equation, not autoparallels of the spacetime connection. One way to see why this is the case is starting from the action for free particles (including free of gravity, which in $f(T)$ means torsion-free). The action for such a free particle is given by
\begin{equation}
 S= \bigintsss\, \left[\eta_{\mu\nu} \frac{dx^{\mu}}{d\tau} \frac{dx^{\nu}}{d\tau}\right]^{\frac{1}{2}} d\tau \,, \label{eq:freepartact}
\end{equation}
where $\eta_{\mu\nu}$ is the spacetime coordinate-frame metric in the absence of gravity:
\begin{equation}
 \eta_{\mu\nu}= \eta_{\hat{\alpha}\hat{\beta}}e^{\hat{\alpha}}_{\;\mu}e^{\hat{\beta}}_{\;\nu}\,, \label{eq:nogravmet}
\end{equation}
with $e^{\cdot}_{\;\cdot}$ the gravity-free orthonormal tetrads which project from the orthonormal frame to the coordinate frame. The gravitational coupling prescription in teleparallel gravity amounts to the replacement of the gravity-free tetrads with the tetrad compatible when torsion is present
\begin{equation}
 e^{\hat{\alpha}}_{\;\mu} \rightarrow h^{\hat{\alpha}}_{\;\mu}\,,
\end{equation}
so that in the presence of torsion (\ref{eq:nogravmet}) becomes
\begin{equation}
  g_{\mu\nu}= g_{\hat{\alpha}\hat{\beta}}h^{\hat{\alpha}}_{\;\mu}h^{\hat{\beta}}_{\;\nu}\,, \label{eq:gravmet}
\end{equation}
($\eta_{\hat{\alpha}\hat{\beta}}$ and $g_{\hat{\alpha}\hat{\beta}}$ are of course numerically equivalent, but differ conceptually \cite{ref:dasdeb} and most authors do not distinguish).

It can be seen that employing this coupling principle essentially replaces the gravity-free metric in the action (\ref{eq:freepartact}) with the gravitational metric so that (\ref{eq:freepartact}) becomes
\begin{equation}
 S= \bigintsss\, \left[g_{\mu\nu} \frac{dx^{\mu}}{d\tau} \frac{dx^{\nu}}{d\tau}\right]^{\frac{1}{2}} d\tau \,. \label{eq:partact}
\end{equation}
As is well-known, extremizing this action with respect to the particle's position and velocity yields the geodesic equation, whose connection is the Christoffel connection:
\begin{equation}
\frac{d^{2}x^{\alpha}}{d\tau^{2}}_{|x^{.}=\chi^{.}(\tau)}=-\left(\mathring{\Gamma}^{\alpha}_{\mu\nu} \frac{dx^{\mu}}{d\tau}\frac{dx^{\nu}}{d\tau}\right)_{|x^{.}=\chi^{.}(\tau)}\,, \label{eq:geoeqn}
\end{equation}
where $\rchi^{\cdot}(\tau)$ denotes a restriction to the parameterized geodesic path of the particle. A fuller account of how this gravitational coupling prescription arises may be found in \cite{ref:booktor}.

Since the free-falling particle motion is geodesic, the geodesic deviation equation applies to free-falling particles exactly like in curvature-only theories. That is,
\begin{equation}
 \mathring{\nabla}_{\mathbf{u}}\mathring{\nabla}_{\mathbf{u}} \xi^{\alpha}=\mathring{R}^{\alpha}_{\;\;\mu\nu\beta} u^{\mu}u^{\nu}\xi^{\beta}\,, \label{eqLgeodev} 
\end{equation}
where $u^{\mu}$ are the components of $\mathbf{u}$, which is tangent to the geodesics, and $\xi^{\alpha}$ the deviation vector. Therefore, pathologies in the Riemann-Christoffel tensor in $f(T)$ gravity herald a pathology in the tidal forces on free particles, just as in general relativity or similar curvature-based theories. Specifically we should consider this tensor in some orthonormal frame since in geodesic deviation the tensor is projected onto 4-velocities. (The orthonormal components also eliminate spurious coordinate artifacts, which could be a false signal of a singularity.) Alternatively, if we are willing to lose some information, we can consider its Kretschmann scalar.


\section[The spherically symmetric vacuum]{The spherically symmetric $\bm{f(T)}$ vacuum}\label{sec:sphvac}

In this work we will be considering an action of the form (\ref{eq:tegraction}) with $f(T)$ specifically given by
\begin{equation}
 f(T)=T +\frac{\alpha}{2}T^{2}\,. \label{eq:ourlag}
\end{equation}
This form of the action is considered important for several reasons. One is that it is the torsion analog of Starobinsky theory \cite{ref:power1} - \cite{ref:power5}, and hence many of the arguments in favor of Starobinsky theory in the curvature realm could apply to this theory in the arena of torsion theories. Also, if the full $f(T)$ Lagrangian density function is considered to be one analytic in $T$ then (\ref{eq:ourlag}) yields the lowest-order correction beyond TEGR. In this manuscript however we make no claim to the smallness of the quadratic term, and study the exact (as opposed to perturbative) properties of vacuum solutions. Perturbative torsion vacuum solutions have been discussed in \cite{ref:solar1} - \cite{ref:solarend}. In covariant $f(T)$ theory perturbative solutions were discovered in \cite{ref:ourvac} and were further studied in \cite{ref:bahamondevac},  \cite{ref:golovg} ,\cite{ref:pfeifer}, \cite{ref:defang}.

Here we will consider the physically relevant scenario of static spherical symmetry specifically in the \emph{isotropic coordinate chart}, for reasons which will be discussed  below. A  line element compatible with such a chart is given by:
\begin{equation}
\Scale[0.90]{ ds^{2}=A^{2}(\rho)\,dt^{2} - B^{2}(\rho)\left[d\rho^{2} +\rho^{2}\left(d\theta^{2} + \sin^{2}\theta\, d\phi^{2}\right)\right]\,,}
\label{eq:isomet}
\end{equation}
with $t_{1}< t < t_{2}$, $\rhoh \leq \rho < \infty$, $0 < \theta < \pi$, $0\leq \phi < 2\pi$. In (\ref{eq:isomet}) a horizon exists where $A(\rho)=0$ and the value of $\rho$ where this occurs will be denoted as $\rhoh$. 

A metric compatible tetrad in this coordinate system is provided by
\begin{equation} \small
\Scale[0.90]{\left[h^{\hat{\alpha}}_{\;\;\mu}\right]=\left[ \begin{array}{cccc}
\xi_{(0)}A(\rho) & 0 & 0 & 0\\
0 & \xi_{(1)}B(\rho) & 0 & 0 \\
0 & 0 & \xi_{(2)}B(\rho)\rho & 0 \\
0 & 0 & 0 & \xi_{(3)}B(\rho)\rho\sin\theta  \end{array} \right]\,,} \label{eq:diagtet}
\end{equation}
where the $\xi_{(\mu)}$ can each be either $+1$ or $-1$ independently of each other.
We note that this coordinate system is not suitable for describing the region interior to the horizon and therefore we restrict our analyses to horizons and their exterior regions.

Before continuing we must compute the spin connection to ensure that we are studying Lorentz covariant $f(T)$ gravity. 
We utilize both of the previously mentioned methods to compute the inertial spin connection components. In the case here tetrad (\ref{eq:diagtet}) is used, and the flat-space (or $G\rightarrow 0$) limit described earlier corresponds to taking $A(\rho)=1$ and $B(\rho)=1$, and their derivatives set to zero. 

In the case of tetrad (\ref{eq:diagtet}) both methods yield the same spin connection components, as they should. These components are:
\begin{align}
\omega^{\hat{\rho}\hat{\theta}}{}_\theta = - \omega^{\hat{\theta}\hat{\rho}}{}_\theta = \frac{\xi_{(2)}}{\xi_{(1)}}, \nonumber \\
\omega^{\hat{\rho}\hat{\phi}}{}_\phi = - \omega^{\hat{\phi}\hat{\rho}}{}_\phi = \frac{\xi_{(3)}}{\xi_{(1)}}\sin\theta, \nonumber \\
\omega^{\hat{\theta}\hat{\phi}}{}_\phi = - \omega^{\hat{\phi}\hat{\theta}}{}_\phi = \frac{\xi_{(3)}}{\xi_{(2)}}\cos\theta
\,. \label{eq:ourspincon}
\end{align}
These spin connection components turn out to be similar to the ones one would get if the more common Schwarzschild coordinates were used instead of isotropic coordinates. The equations of motion that result with the tetrad (\ref{eq:diagtet}) and spin connection (\ref{eq:ourspincon}) do not depend on whether or not any combination of the $\xi_{(\mu)}$ are $+1$ or $-1$, indicating time-reversal, parity and rotational invariance as required by full Lorentz symmetry. (There is local boost invariance as well.) Changing the the sign of only some of the spatial $\xi_{(\mu)}$ is equivalent to either a rotation or a parity transformation plus a specific rotation. Only even powers of the $\xi_{(\mu)}$ appear in the resulting equations of motion. Since the signs of the $\xi_{(\mu)}$ are irrelevant, from this point onward we will set all $\xi_{(\mu)} =+1$ without loss of generality. 

It can be easily confirmed that now the resulting theory is Lorentz covariant. For example, one could take the tetrad (\ref{eq:diagtet}) and apply a local (coordinate dependent) proper Lorentz transformation, $\Lambda_{\hat{\alpha}}^{\;\hat{{\beta}^{\prime}}}(x)$ to it,
\begin{equation}
\Lambda_{\hat{\alpha}}^{\;\hat{{\beta}^{\prime}}}(x)h^{\hat{\alpha}}_{\;\;\mu} =  h^{\hat{\beta}^{\prime}}_{\;\;\mu}\,. \label{eq:lttet}
\end{equation}
Then it can be verified that the action calculated with $T$, constructed from $h^{\hat{\alpha}}_{\;\;\mu}$ via using (\ref{eq:diagtet}) and (\ref{eq:ourspincon}) in (\ref{eq:genT}), is exactly the same as the action computed with $T$ constructed from $h^{\hat{\beta}^{\prime}}_{\;\;\mu}$ and the corresponding spin-connection of the method of \cite{ref:KandS} and \cite{ref:lomonosov}. Said another way, the torsion scalar $T$ transforms as a scalar under local Lorentz transformations (including parity and time-reversal) if one includes the proper inertial spin connection (which, for the scenarios studied here yields a torsion-free Minkowski limit as discussed previously), and also the resulting equations of motion from that action are locally Lorentz invariant. 

In the case of this paper, the explicit form of the Lorentz invariant torsion scalar, using (\ref{eq:diagtet}) and (\ref{eq:ourspincon}) in (\ref{eq:gentor}) and computing (\ref{eq:genT}), is
\begin{equation}
 T = \frac{2B'(2BA'+AB')}{AB^4}\,, \label{eq:ourT}
\end{equation}
stressing that this is in isotropic coordinates, and hence does not exactly resemble the covariant torsion scalar in the usual curvature coordinates.
In (\ref{eq:ourT}) the primes denote differentiation with respect to $\rho$ and we have suppressed the explicit $\rho$ dependence of the tetrad functions.

The vacuum equations of motion, by using (\ref{eq:diagtet}) and (\ref{eq:ourspincon}) in (\ref{eq:eoms}) are given by
\begin{subequations}\allowdisplaybreaks
\begin{align}
L_{\hat t \hat t} 
  =& \frac{-2 B \rho B''+\rho \left(B'\right)^2-4 B B'}{B^4 \rho } \nonumber \\
  & + \frac{\alpha}{A^2 B^8 \rho } \Big[
      B' \left(A B' \left(-4 B \rho \left(2 B A''+3 A B''\right)\right.\right. \nonumber \\
      &+17 A\rho
      \left.\left(B'\right)^2-8 A B B'\right) \nonumber \\
  & -8 A B A' \left(2 B \rho B''-3 \rho 
      \left(B'\right)^2+2 B B'\right) \nonumber \\
      &  \left.+4 B^2 \rho \left(A'\right)^2 B'\right) \Big] = 0\,, \label{eq:eomtt} \\
L_{\hat \rho \hat \rho } 
  =& \frac{2 B A' \left( \rho  B'+B\right)
  + A B' \left( \rho  B'+2 B\right)}{A B^4 \rho } \nonumber \\
  & + \frac{\alpha}{A^2 B^8 \rho } \Big[
    B' \left(2 B A'+A B'\right) \left(2 B A' \left(3 \rho B' \right.\right.  \nonumber \\
    & \left. +2 B\right)+AB' \left.\left(3 \rho B'+4 B\right) \right) \Big] =0\,, \label{eq:eomrhorho} \\
L_{\hat \vartheta \hat \vartheta} 
  =&  \frac{B^2 \left( \rho  A''+A'\right)
  - A \rho \left(B'\right)^2+A B \left( \rho  B''+B'\right)}{A B^4 \rho } \nonumber \\
  & + \frac{\alpha}{A^3 B^8 \rho } \Big[
  A^2 \left(B'\right)^2
  \left(6 B \rho \left(B A''+A B''\right)\right. \nonumber \\
  &-9 A \rho \left(B'\right)^2+2 A B
  \left. B'\right)- 4 B^3 \rho\left(A'\right)^3 B'  \nonumber \\
  &+4 A B^2 \left(A'\right)^2 \left(B \rho
  B''+B' \left(B-3 \rho B'\right)\right) \nonumber \\
  & + 2 A B A' B' \left(2 B \rho \left(2 B A''+3
  A B''\right)\right. \nonumber \\
  &-\left. 10 A \rho \left(B'\right)^2+3 A B B'\right)
  \Big]=0 \, . \label{eq:eomangang}
\end{align}
\end{subequations}

In the following sections we study several relevant properties of vacuum solutions. In section \ref{sec:center} we study properties that vacuum solutions should possess in order to be regular at their center. We assume in that section that there are no horizons so that $\rho=0$ can validly be covered by the coordinate chart of (\ref{eq:diagtet}). In general relativity (or TEGR), this requirement of regularity of the vacuum at the origin of course leads to the well-known conclusion that the vacuum spacetime must be Minkowski spacetime everywhere. (In general relativity Minkowski spacetime is the only spherically symmetric \emph{everywhere vacuum} solution that is regular everywhere.)  We reiterate that in this manuscript we are solely concerned with vacuum solutions. It is not immediately clear though that this restriction to Minkowski spacetime remains in more general extended $f(T)$ settings since the equations are no longer Einstein vacuum equations and, as well, it is arguably less obvious what is meant by ``regularity'' in a Weitzenb\"{o}ck spacetime. Of course though, Minkowski spacetime is indeed a vacuum solution of $f(T)$ gravity, but it is not clear that it must be the \emph{only} spherically symmetric vacuum solution with a regular center in $f(T)$ gravity since from (\ref{eq:eoms}) $f(T)$ vacuum gravity can mimic general relativity with a peculiar matter source.

In section \ref{sec:horizons} we assume that there is a horizon somewhere in the spacetime, and study the properties that solutions must have in order for the horizon's existence. In section \ref{sec:computational} we supplement the analytic work with numerical evolutions and also summarize the various scenarios. Finally, some comments are made regarding the presence of a cosmological constant.


\section[Properties near the center]{Properties near the center}\label{sec:center}

In this section we will study the properties that vacuum solutions must possess near $\rho=0$ subject to the condition that the solutions are regular. We assume in this section that there are no horizons so that the isotropic coordinate chart can cover $\rho=0$.

To begin the study of regular centers let us start with studying the geodesic equations (\ref{eq:geoeqn}). Constructing the Christoffel connection with (\ref{eq:isomet}) and concentrating on radial geodesics (i.e. only $u^{t}$ and $u^{\rho}$ not being zero), the $\rho$ component of equation (\ref{eq:geoeqn}) yields:
\begin{equation}\small
  \frac{d^{2}\rho}{d\tau^{2}}_{|x^{.}=\chi^{.}(\tau)}= - \left[\frac{AA^{\prime}}{B^{2}} \left(u^{t}\right)^{2} + \frac{B^{\prime}}{B}\left(u^{\rho}\right)^{2}\right]_{|x^{\cdot}=\chi^{\cdot}(\tau)}    \label{eq:rgeo}
\end{equation}
Put in the language of ``effective force'', one can look at equation (\ref{eq:rgeo}) as a force equation where the left-hand side is the effective acceleration of the particle in the $\rho$ direction, and the right-hand side represents the force. In fact, viewing the geodesic equation as a force equation is actually the correct interpretation in teleparallel gravity since gravitation in $f(T)$ theory is considered a true force.

Consider now placing a massive test particle at $\rho=0$ with \emph{no} initial spatial velocity. Spherical symmetry dictates that no radial direction from the center is privileged due to isotropy about this point, so the particle, having zero initial spatial velocity, should not start to move away from the origin.  In other words, the left-hand side of (\ref{eq:rgeo}) must be zero for such a particle. Now, from this argument we know that for such a particle at the center we must have $u^{\rho}=0$ for all $\tau$. The only way that the right-hand side of (\ref{eq:rgeo}) can vanish for $u^{t}\neq 0$ and $B$ not infinite is if $A^{\prime}=0$. $A(0)=0$ is not considered as it would indicate an infinite redshift horizon at the center and in this part of the paper horizons are not considered.

Next consider a radially in-falling particle as it crosses $\rho=0$. Here as well when the particle is momentarily at the center it should not be pulled in any direction due to spherical symmetry, so again both sides of (\ref{eq:rgeo}) must be zero at $\rho=0$. In this case however neither $u^{t}$ nor $u^{\rho}$ are zero at the center. We have just argued above though that at $\rho=0$ the derivative of $A$ must vanish.  Assuming that the metric function $B$ is not infinite, this implies that at the center $B^{\prime}(0)=0$ in order to make the right-hand side of (\ref{eq:rgeo}) vanish. These arguments do not explicitly rely on the spacetime being vacuum, and so could apply also, for example, at the centers of spherical stars.

Before continuing we comment that it is perhaps interesting to note that, assuming that if neither of the tetrad functions or their first two derivatives are infinite, the equations of motion at places where $A^{\prime}$ and $B^{\prime}$ simultaneously vanish are locally equivalent to Einstein's equations. This can be seen from (\ref{eq:eomtt})-(\ref{eq:eomangang}) where under these conditions the terms proportional to $\alpha$ locally vanish.

In summary, we find that for a regular center to exist, regular meaning acceptable particle-motion, both $A^{\prime}(0)$ and $B^{\prime}(0)$ must be zero. This is similar to the conditions in general relativity where a kink in the metric at the origin is forbidden as it would imply the presence of an infinitely thin segment of matter there.

Our analysis above is local. However, \emph{if} the spacetime is everywhere vacuum and regular, then there is arguably no preferred center of symmetry, and $\rho=0$ could be placed anywhere in the spatial submanifold of the spacetime. The above arguments would then apply to every point in the spacetime and we would conclude that anywhere in the globally vacuum spacetime the first derivatives of $A$ and $B$ should vanish. This would imply that the spacetime is Minkowski spacetime everywhere, and so even in $f(T)$ gravity Minkowski spacetime remains the globally vacuum spherically symmetric solution that is everywhere regular. This result is perhaps not surprising since in such a spacetime we have promoted isotropy about a point to isotropy about every point. It is important to stress that when we refer to vacuum in this manuscript we are referring to the absence of a cosmological constant as well. Some comments on the cosmological constant will be made in section \ref{sec:cosconst}.

We can analyze this claim of global Minkowski structure more quantitatively by assuming that the functions $A$ and $B$ are analytic functions. We can therefore Taylor expand the equations around $\rho=0$, assuming $A^{\prime}(0)$ and $B^{\prime}(0)$ are zero, as dictated by the above analysis of the geodesic equations. Since we are solving the vacuum equations in some non-zero domain around $\rho=0$, the resulting equations must equal zero order-by-order.

We begin by analyzing equation (\ref{eq:eomtt})
and expanding it about $\rho=0$
subject to the above mentioned regularity condition
that $A'(0) = 0 = B'(0)$.
The lowest order term in $L_{\hat t \hat t}$ implies that
\begin{equation} \label{eq:centereomtt}
- \frac{6 B''(0)}{B^3(0)} + \mathcal{O} (\rho) = 0 \, .
\end{equation}
This equation may be satisfied by demanding that $B''(0) = 0$.
Next the equation of motion (\ref{eq:eomrhorho})
is analyzed to lowest non-trivial order
and once $B''(0) = 0$ is employed it demands that
\begin{equation} \label{eq:centereomrhorho}
\frac{2 A''(0)}{ A(0) B^2(0) } + \mathcal{O} (\rho) = 0 \, .
\end{equation}
This implies that $A''(0) = 0$.
We then repeat the above procedure order-by-order
of first analyzing equation (\ref{eq:eomtt}),
which must be satisfied by setting
the next higher derivative of $A$ to zero.
We then use this condition in the equation of motion (\ref{eq:eomrhorho})
which tells us that the next higher derivative of $B$ must also equal zero.
The pattern can be seen to arise
as far as the expansion could be carried out.
Therefore, it is conjectured that
all-order derivatives of $A$ and $B$ must vanish
when demanding regularity in the sense
required by the equation (\ref{eq:rgeo}).
This implies that the vacuum spacetime is Minkowski
since in this particular argument the so called
flat-functions are ruled out by our assumption of real analyticity.
We confirm that the resulting spacetime is indeed Minkowski
in Sec.~\ref{sec:computational} (see Fig.~\ref{fig:numtests}).
There we evolve solutions from $\rho=0$
subject to the condition that $A^{\prime}(0)=0$ and $B^{\prime}(0)=0$
and always obtain Minkowski spacetime throughout the entire domain.

Before proceeding to the next section on horizons we summarize the findings of this section as follows: Requiring that the gravitational force equation, which in $f(T)$ gravity is equivalent to the geodesic equation, behaves properly at the center of symmetry dictates that $A^{\prime}(0)=0$ and $B^{\prime}(0)=0$. One may then argue that if the spacetime is vacuum everywhere and further one demands the spacetime to be geodesically regular everywhere, then $\rho=0$, the center of symmetry, may be taken to be anywhere in the spatial submanifold. This then implies if the spherically symmetric spacetime is vacuum and regular everywhere all spatial positions should have the condition of vanishing first derivatives, and hence the spacetime is Minkowski spacetime. This statement was then quantified by assuming that the tetrad is an analytic one, and solving the field equations order-by-order about $\rho=0$ indeed yields Minkowski spacetime. 

On the other hand, a vacuum spacetime with a horizon does not need to be regular everywhere in order to be physically acceptable if the singular point is hidden behind a horizon. This is the hypothesis of cosmic censorship.  Therefore the above results do not necessarily apply to vacuum spacetimes with horizons as we  do not need to worry about regularity inside a horizon and therefore relax the condition of globally Minkowski spacetime. We will analyze situations with horizons in the next section.


\section[Horizon analysis]{Horizon analysis}\label{sec:horizons}
In this section we are interested in the properties at possible horizons, which by definition occur at $\rho=\rhoh$ where $A(\rhoh)=0$. First we will make some general analysis regarding horizons and the equations of motion (\ref{eq:eomtt} - \ref{eq:eomangang}). It is assumed that $B(\rhoh)$ is not infinite, since in the isotropic coordinates this would imply that the horizon has infinite proper area. It is also assumed that $B(\rhoh)$ is not zero as that would yield a horizon of zero proper area.

We begin by writing each individual equation of motion over a common denominator. It is then noted that where a possible horizon occurs ($A(\rhoh)=0$) the resulting common denominators all vanish, and therefore if the equations of motion are to be equal to zero we must have that the numerators must vanish there. The resulting numerators subject to the $A=0$ horizon condition are as follows:
\begin{subequations}
\begin{align}
4 \alpha r B^2 A'^2 B'^2 & = 0 \label{eq:numertt} \\
4 \alpha B^2 A'^2 B' ( 2B + 3rB' ) & = 0 \label{eq:numerrhorho} \\
4 \alpha r B^3 A'^3 B' & = 0 \,. \label{eq:numerangang}
\end{align}
\end{subequations}
Under the specified conditions,
the only way that the above equations can be zero at $\rhoh$
is if the following condition holds:
\begin{equation}
 \left(A^{\prime}(\rho)B^{\prime}(\rho)\right)_{|\rho=\rhoh}=0\,. \label{eq:horizoneomcond}
\end{equation}
We note that in TEGR the above restriction is not required
since $\alpha = 0$ for TEGR and so (\ref{eq:numertt}-c) are identically zero for TEGR.

Regarding condition (\ref{eq:horizoneomcond}),
we will next show that it is specifically $A^{\prime}(\rho)$
that should vanish at $\rhoh$
in order for a regular horizon solution to possibly exist in the sense of no Riemann-Christoffel singularity. In other words, it will be shown that if $A^{\prime}(\rhoh)=0$ there may potentially be a non-singular horizon solution, and that solution will be regular subject to the further restriction that $A^{\prime\prime}(\rhoh)=0$. As we will discuss, this last condition is not practically achieved with an asymptotically Schwarzschild tetrad when $\alpha\neq 0$.  

Following that analysis, it will separately be shown that, under the assumption of analyticity, it is $B^{\prime}(\rhoh)$
that should vanish for the equations to possibly possess a horizon solution (regular or otherwise). It will also be shown though that such a condition cannot solve the equations of motion to all orders, so actually there is no solution under this condition. This criterion is derived subject to the condition of analyticity and so the analysis there is restricted to analytic tetrad solutions. This is why both the Riemann-Christoffel analysis and the analysis involving Taylor expansions of the equations of motion will both be utilized. The two conditions do not encompass each other. Failure of the former condition allows for a possible horizon solution to the equations of motion, but the horizon is Riemann-Christoffel singular. Failure of the second condition disallows any horizon solution to the equations of motion, but may not necessarily apply to non-analytic tetrads.

Let us now begin the first of the above mentioned analyses by examining the orthonormal components of the Riemann-Christoffel tensor in a frame locally adapted to the coordinate system. Specifically we examine the component
\begin{equation}
 \mathring{R}_{\hat{\theta}\hat{t}\hat{t}\hat{\theta}}= \frac{A^{\prime}\left(B + \rho B^{\prime}\right)}{\rho A B^{3}} \,. \label{eq:riemangt}
\end{equation}
Recall that on a horizon $A(\rhoh)=0$, and that we are not considering $B(\rhoh)$ zero or infinite. Since we have determined above that a non-singular horizon requires $A^{\prime}(\rhoh)B^{\prime}(\rhoh)=0$, (\ref{eq:riemangt}) can only possibly be regular if $A^{\prime}(\rhoh)=0$. Next let us concentrate on the Riemann component
\begin{equation}
 \mathring{R}_{\hat{t}\hat{\rho}\hat{t}\hat{\rho}}=\, \frac{A^{\prime}B^{\prime}- A^{\prime\prime}B}{AB^{3}} \, \label{eq:riemrt}
\end{equation}
at $\rho=\rhoh$, now subject to the conditions $A(\rhoh)=0$ and $A^{\prime}(\rhoh)=0$. We can see that in order for (\ref{eq:riemrt}) to not be infinite we now also require the condition $A^{\prime\prime}(\rhoh)=0$. 

At this stage it has been established that for a regular horizon in the orthonormal Riemann-Christoffel sense when $\alpha \neq 0$, the conditions required are that the tetrad function $A$ as well as its first two derivatives must vanish. It will be shown below in section \ref{sec:computational} with numerical work, that for asymptotically Schwarzschild spacetimes, the second derivative of $A$ at the possible horizon is not zero. Therefore a Riemann-Christoffel curvature singularity will exist there if a horizon forms. 

The above findings rely on using a result from computational evolutions in order to show that a regular horizon does not exist; namely the result that $A^{\prime\prime}$ does not equal to zero on the horizon which violates the studied regularity condition there. We also repeat that the numerical results stem from evolving asymptotically Schwarzschild black holes, which are arguably the most physically relevant in this paradigm. It would though be beneficial if a no-go argument for regular horizons could be implemented without relying on a numerical result, and that is also purely local. To accomplish this we will do a similar analysis to what was done earlier. That is, we will expand the equations of motion about $\rho=\rhoh$ subject to the condition that $A(\rhoh)=0$, making the assumption that the functions $A(\rho)$ and $B(\rho)$ are analytic in their non-zero domain of convergence of $\rho \geq \rhoh$.  (The function $B(\rho)$ could in principle be considered formally Laurent expandable, removing its requirement for analyticity near the horizon, but we shall consider it only Taylor expandable since $B(\rhoh)$ becoming infinite implies a horizon of infinite proper size.) 

The isotropic coordinates (\ref{eq:isomet}), (\ref{eq:diagtet}) are chosen in this paper due to the fact that they lend themselves better to the study of horizons in terms of analytic tetrad functions. The famous Schwarzschild black hole of general relativity (or TEGR), for example, when cast in isotropic coordinates, can be described by an analytic diagonal tetrad such as (\ref{eq:diagtet}):
\begin{subequations}\allowdisplaybreaks
\begin{align}
&A_{\mbox{\tiny{Schw}}}(\rho)=\frac{\rho-{M}/{2}}{\rho+{M}/{2}}\nonumber \\
&=\frac{1}{M}\left(\rho-\rhoh\right) -\frac{1}{M^{2}}\left(\rho-\rhoh\right)^{2} +\mathcal{O}\left(\rho-\rhoh \right)^{3}\,, \label{eq:schwaexp} \\
&B_{\mbox{\tiny{Schw}}}(\rho)=\left(\frac{\rho+{M}/{2}}{\rho}\right)^{2} \nonumber \\
&= 4 -\frac{8}{M}\left(\rho-\rhoh\right) +\frac{20}{M^{2}}\left(\rho-\rhoh\right)^{2} +\mathcal{O}\left(\rho-\rhoh\right)^{3}\,. \label{eq:schwbexp}
\end{align}
\end{subequations}
The Schwarzschild horizon is located at $\rhoh=M/2$ and it is explicitly assumed that $M >0$. Isotropic coordinates are also somewhat better behaved at the horizon than the usual Schwarzschild coordinates. For example, the most often used characters of the Riemann-Christoffel tensor in the coordinate frame, $\mathring{R}_{\alpha\beta\gamma\delta}$ or $\mathring{R}^{\alpha}_{\;\;\beta\gamma\delta}$, are all finite at $\rho=\rhoh$ for the Schwarzschild metric in isotropic coordinates, whereas in Schwarzschild coordinates at least some components diverge on the Schwarzschild horizon\footnote{Of course, in the orthonormal frame all components are finite at the Schwarzschild horizon, indicating that it is not a true curvature singularity.}. The importance of the Riemann-Christoffel tensor in $f(T)$ gravity, whose connection is instead the flat Weitzenb\"{o}ck connection, was discussed above.

We begin with an expansion about some value $\rho=\rhoh$ where the condition $A(\rhoh)=0$ is assumed to hold. That is, we begin with the assumption that a horizon exists somewhere. The series expansion to leading order yields
\begin{subequations}\allowdisplaybreaks
\begin{align}
L_{\hat{t}\hat{t}}=& \frac{4\alpha B^{\prime 2}(\rhoh)}{B^{6}(\rhoh)} \frac{1}{(\rho-\rhoh)^{2}} + \mathcal{O}(\rho-\rhoh)^{-1}\,, \label{eq:lowtt} \\
L_{\hat{\rho}\hat{\rho}}=& \frac{4\alpha B^{\prime}(\rhoh)\left(2B(\rhoh)+3\rhoh B^{\prime}(\rhoh)\right)}{\rhoh B^{6}(\rhoh)(\rho-\rhoh)^{2}}  \nonumber \\
 & + \mathcal{O}(\rho-\rhoh)^{-1}\,, \label{eq:lowrhorho} \\
L_{\hat{\theta}\hat{\theta}}=& \frac{4\alpha B^{\prime}(\rhoh)}{B^{5}(\rhoh)} \frac{1}{(\rho-\rhoh)^{3}} + \mathcal{O}(\rho-\rhoh)^{-2}\,. \label{eq:lowangang}
\end{align}
\end{subequations}
Interestingly we note that the potentially most singular terms in the equations of motion are proportional to $\alpha$, hinting that perhaps TEGR tends to be less singular than extended teleparallel gravity in this setting. 

From the above equations we note that for all three equations to be equal to zero, the physically acceptable condition that $B^{\prime}(\rhoh)=0$ must be employed. This condition is not required in TEGR since the above terms are automatically zero when $\alpha=0$. The case $B(\rhoh)$ being infinite is not considered due to it representing a horizon of infinite size, and also violates the assumption that $B$ is an analytic function in the neighborhood of $\rhoh$. 

Next we continue to analyze the equations of motion now subject to the conditions $A(\rhoh)=0$ and the newly discovered condition that $B^{\prime}(\rhoh)=0$. This yields, to leading order,
\begin{subequations}
\allowdisplaybreaks{\begin{align}
L_{\hat{t}\hat{t}}=&- \frac{2 B^{\prime\prime}(\rhoh) \left(B^{3}(\rhoh)+6 \alpha B^{\prime\prime}(\rhoh)\right)}{B^{6}(\rhoh)} \nonumber \\
&+ \mathcal{O}(\rho-\rhoh) \,, \label{eq:low2tt} \\
L_{\hat{\rho}\hat{\rho}}=& \frac{2 \left(B^{3}(\rhoh)+4 \alpha B^{\prime\prime}(\rhoh)\right)}{\rhoh B^{5}(\rhoh)(\rho-\rhoh)} + \mathcal{O}(\rho-\rhoh) \,, \label{eq:low2rhorho} 
\end{align}}
\end{subequations}
where $L_{\hat{\theta}\hat{\theta}}$ has not been written as it will not be required for the argument. We now note that equations (\ref{eq:low2tt}) and (\ref{eq:low2rhorho}) cannot \emph{both} be set equal to zero under a physically acceptable condition. Therefore, we conclude that if $A(\rhoh)=0$ we cannot simultaneously solve all vacuum equations of motion outside the horizon.

We mention again
that the analysis on equations (\ref{eq:low2tt}) and (\ref{eq:low2rhorho})
does not apply to TEGR since (\ref{eq:low2tt}) and (\ref{eq:low2rhorho})
are subject to the $B'(\rhoh)=0$ condition,
which as stated previously is not a requirement of TEGR.
This can easily be seen from equations (\ref{eq:lowtt}) and (\ref{eq:lowrhorho})
where the leading term vanishes automatically with $\alpha=0$.
(The Schwarzschild horizon has the value $B^{\prime}(\rhoh)= -8/M$.)

Although the above mentioned analyticity restriction is mild
and certainly applies to the analogous black hole in general relativity/TEGR,
as shown in (\ref{eq:schwaexp},b),
it does impose a limitation on the applicability of the analysis to the set of analytic tetrads. Therefore in section \ref{sec:computational} we study solutions numerically which are asymptotically Schwarzschild. Since in the weak field the Schwarzschild metric is known to give excellent agreement with observations \cite{ref:schwsolbook} it is expected that far from the horizon the spacetime metric mimics closely the Schwarzschild one. In this current section however we made no assumptions about the asymptotics far away from the horizon. 

The findings in this section can be summarized as follows: The equations of motion can potentially have a solution when $A(\rhoh)=0$ subject to condition (\ref{eq:horizoneomcond}). If one then wishes that solution to be non-singular in the Riemann-Christoffel sense, the supplementary conditions $A^{\prime}(\rhoh)=0$ and as well  $A^{\prime\prime}(\rhoh)=0$ arise. Numerically it will be found that if a horizon exists this last condition does not hold. However, if one is willing to restrict the analysis to analytic functions one does not need to rely on the numerical result. Then one finds that in a scenario where $A(\rhoh)=0$ the equations of motion do not have a solution in the vicinity of the horizon.


\section[Computational results]{Computational results}\label{sec:computational}

Here we will construct numerical solutions
to the static spherically symmetric vacuum equations of motion
of the $f(T) = T + \frac \alpha 2 T^2$ gravity theory.
The analysis is constrained to the solutions
that inherit the asymptotic behaviour of the Schwarzschild solution
at spatial infinity and are therefore of potential astrophysical interest.
The Schwarzschild solution will be used
to provide us with the initial data near space-like infinity,
needed for inbound numerical integrations of equations of motion.
We would like to also verify
that the perturbative asymptotically Schwarzschild solution
found in \cite{ref:ourvac} is indeed valid in the weak field regime.
We will therefore next briefly review the perturbative solution.

\subsection{Perturbative solution}

In the isotropic coordinate chart of (\ref{eq:isomet})
the perturbative solution can be written as
\begin{subequations}
\begin{align}
A(\rho) & = A_{\mathrm{Schw}}(\rho) + \alpha a(\rho) \, , \\
B(\rho) & = B_{\mathrm{Schw}}(\rho) + \alpha b(\rho) \, ,
\end{align}
\end{subequations}
where the Schwarzschild functions
$A_{\mathrm{Schw}}$ and $B_{\mathrm{Schw}}$ are given by
(\ref{eq:schwaexp}) and (\ref{eq:schwbexp}),
while functions $a$ and $b$ constitute the perturbative part.
The above expressions are substituted into the equations of motion
(\ref{eq:eomtt})--(\ref{eq:eomangang}),
which are then expanded in powers of the perturbative part.
The leading order terms in this expansion
give a system of differential equations in $a$ and $b$
for which solutions can be obtained in closed form.
Setting the integration constants in these solutions
so that the leading terms in power expansions of $a$ and $B$ about $1/\rho=0$
(i.e.\ spatial infinity) appear at the highest possible order,
$a$ and $b$ follow as
\begin{subequations} \small
\allowdisplaybreaks \begin{align}
a(\rho)
& = \frac{16 M^6 + 70 M^5 \rho + 288 M^4 \rho^2
        + 256 M^3 \rho^3 - 96 M \rho^5} {3M^2 (2\rho+M)^6} \notag \\
& \qquad + \frac{M^2 + 8 M\rho - 4\rho^2}{2M^2 (2\rho+M)^2}
           \ln\frac{2\rho-M}{2\rho+M} \notag \\
& = - \frac{2 M^3}{5 \rho^5} + \mathcal{O} \left(\frac1 \rho\right)^6,
                                                            \label{eq:perta}\\
b(\rho) & = - \frac{M \rho (41 M^3 + 34 M^2 \rho - 12 M \rho^2 - 24 \rho^3)}
          {6 M^2 \rho^2 (2\rho + M)^3} \notag \\
& \qquad + \frac{(2\rho - 3M)(2\rho+M)^4}{8 M^2 \rho^2 (2\rho+M)^3}
                                \ln\frac{2\rho-M}{2\rho+M} \notag \\
& = \frac{2 M^3}{5 \rho^5} + \mathcal{O} \left(\frac 1 \rho \right)^6 .
                                                            \label{eq:pertb}
\end{align}
\end{subequations}

Since the leading terms in the power expansions
of the Schwarzschild solution are
$A_{\mathrm{Schw}}(\rho) = 1 - M/\rho + \mathcal{O} (1/\rho)^2$
and $B_{\mathrm{Schw}}(\rho) = 1 + M/\rho + \mathcal{O} (1/\rho)^2 $,
it is clear the the perturbative solution obeys
the expected asymptotics in this coordinate system. We verify below that this is indeed valid.

\subsection{Numerical procedure}

Standard routines for numerical evolution
of ordinary differential equations will be used.
In order to compactify the semi-infinite radial domain
the dimensionless coordinate defined with $x = \rho / (M+\rho) \in [0,1)$
will be used. Here $M$ is the mass parameter
of the asymptotically Schwarzschild spacetime.
Since we will be using the Schwarzschild solution to provide
us with the initial data for inbound numerical integrations
starting at near spatial infinity,
we will require the Schwarzschild solution in the compactified coordinate.
Using the radial coordinate $x$ defined above
the Schwarzschild solution assumes the form
\begin{subequations}
\begin{align}
A_{\mathrm{Schw}}(x) & = \frac{3x-1}{1+x} \, , \label{eq:schxa} \\
B_{\mathrm{Schw}}(x) & = \frac{(1+x)^2}{4x^2} \, , \label{eq:schxb}
\end{align}
\end{subequations}
where the mass parameter $M$ no longer appears,
and the black hole horizon takes place at $x=1/3$.
In the rescaled equations of motion the only remaining parameter
is the ratio $\alpha/M^2$,
which implies that the asymptotically Schwarzschild solutions
in the $f(T) = T + \frac \alpha 2 T^2$ gravity theory
comprise a one-parameter family of solutions.

As a preliminary test of the reliability of the numerical procedure
we first established that the outbound integration
set off from the vicinity of the center of symmetry
with $A'(0) = B'(0) = 0$ as initial conditions
reproduces the Minkowski spacetime. This will also serve to verify our previous argument that demanding regularity at the center of symmetry of the vacuum spacetime yields Minkowski spacetime globally. 
This test is shown in the upper plot of Fig.~\ref{fig:numtests}
and it confirms our previous analysis in Sec.\ \ref{sec:center} that regularity at the center generates Minkowski spacetime everywhere. 

As a further test, we also verified
that the inbound integration set off from near-infinity
with $\alpha=0$ and with initial conditions
drawn from the Schwarzschild solution
correctly reproduces the Schwarzschild solution
(\ref{eq:schxa}) and (\ref{eq:schxb}) down to the black hole horizon.
This test is shown in the lower plot of Fig.~\ref{fig:numtests},
where one sees only the numerically evolved solution
since it completely overlaps the analytical solution which is also plotted.

\begin{figure}[!h]
\includegraphics[width=\columnwidth]{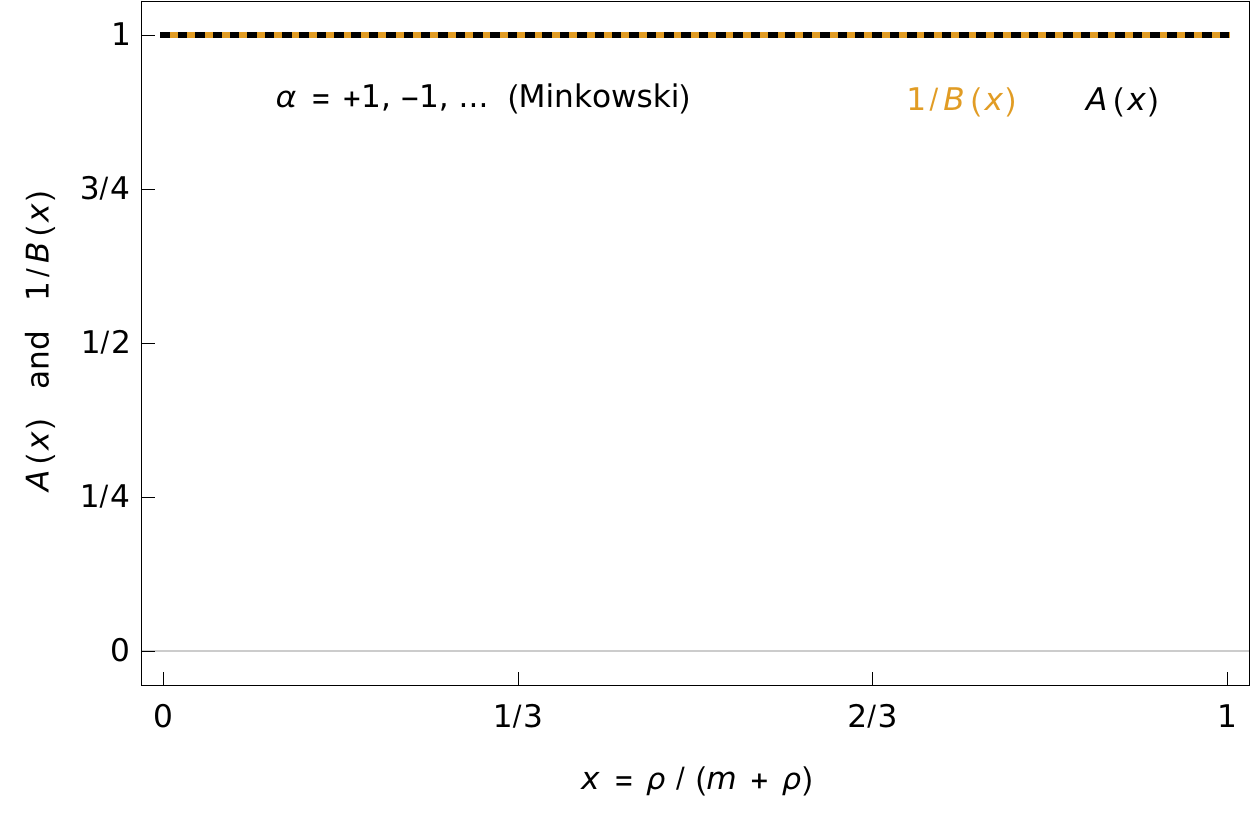} \\[1ex]
\includegraphics[width=\columnwidth]{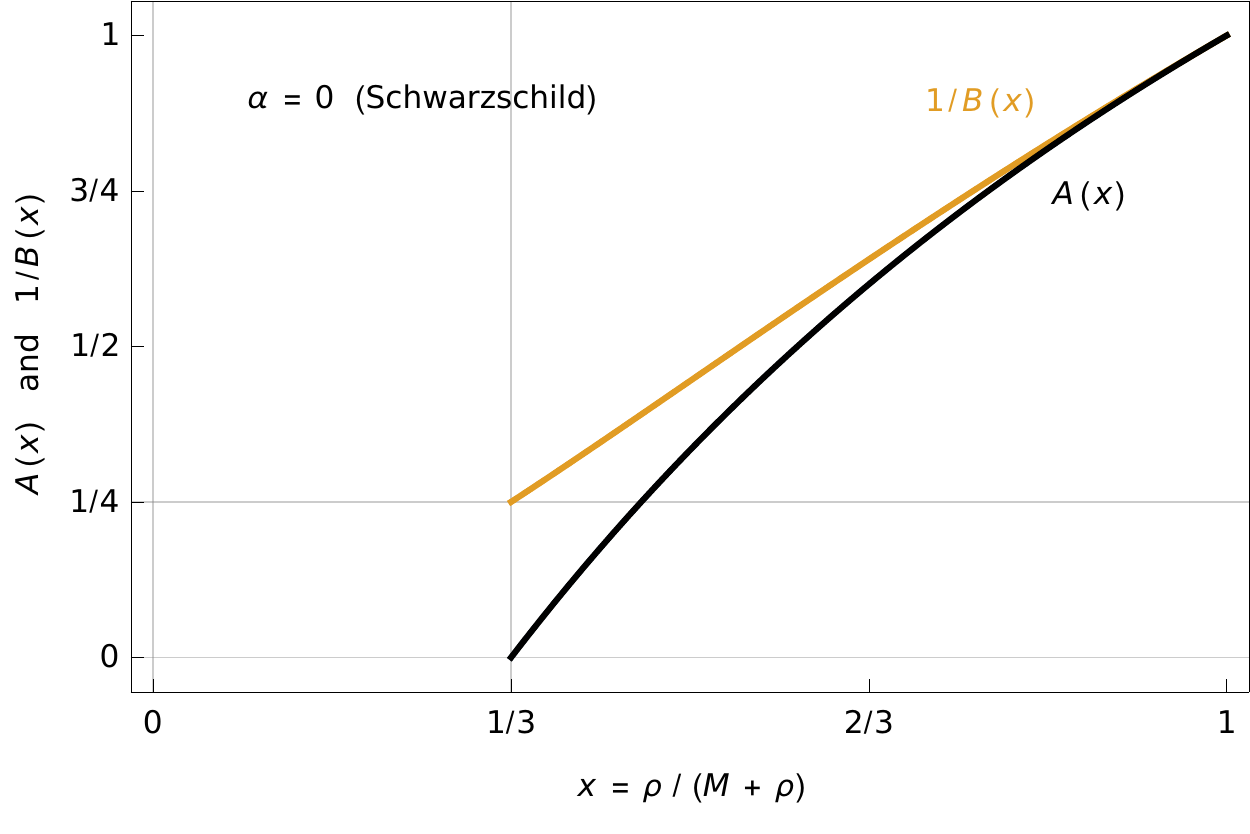}
\vspace*{-5mm}\caption{\label{fig:numtests} \small Tests of the numerical procedure.
Upper plot: Minkowski space is obtained with $\alpha\ne0$
and outbound integration starting from $x=\epsilon$
using $A(\epsilon) = B(\epsilon) = 1$
and $A'(\epsilon) = B'(\epsilon) = 0$ as initial conditions
(solutions $A(x) = 1/B(x) = 1$ overlap).
Lower plot: Schwarzschild exterior solution
functions (\ref{eq:schxa}) and (\ref{eq:schxb})
are reproduced numerically with $\alpha=0$
and inbound integration starting from $x=1-\epsilon$
($A(x)$ is shown in black, $1/B(x)$ in orange). The numerical expressions overlap with the exact expressions indicating a robust computational scheme.}
\end{figure}

As an additional consistency check we compared the
numerical solutions obtained by inbound integrations
from near infinity to the perturbative solutions
(\ref{eq:perta}) and (\ref{eq:pertb}) of \cite{ref:ourvac}
and to the Schwarzschild solution.
As expected, the perturbative solutions
follow the numerical solutions more closely
than the Schwarzschild solution does.
This can be seen in all three plots of Fig.~\ref{fig:numfigs},
where the numerical solutions is shown with thick solid lines,
perturbative solutions is short-dashed,
and the Schwarzschild solution is long-dashed.
The range over which the perturbative and numerical solutions agree
may appear rather short, but this is only an artefact
of the compactification of the radial domain,
and the agreement is actually over a very large region
of the uncompactified domain. In all studies here the validity of the perturbative solutions \cite{ref:ourvac} are confirmed.

\subsection{Numerical results}

In the discussion of the vacuum solutions that
we are about to construct numerically
we will make use of what is sometimes called the GR picture.
Here one interprets the solution originating from a modified
theory of gravity as if it were obtained
within the framework of general relativity,
but with exotic matter being present.
In our case, the exotic matter will be referred to
as the the $f(T)$-fluid.
In the GR picture the vacuum equations of motion can be cast as
$ \mathring G_{\mu\nu} = 8\pi \tilde{\mathcal{T}}_{\mu\nu} $,
where $\tilde{\mathcal{T}}_{\mu\nu}$
is the effective stress-energy tensor of the $f(T)$-fluid.
For general $f(T)$ this effective stress-energy tensor is given by, via (\ref{eq:eoms}),
\begin{equation}
\tilde{\mathcal{T}}_{\mu\nu}  =
\frac{g_{\mu\nu}}{16\pi} \left( T - \frac{f(T)}{f'(T)} \right)
- \frac{S_{\mu\nu}{}^\lambda (\partial_\lambda T) f''(T)}{8\pi f'(T)} \, ,
\end{equation}
while specifically for $f(T) = T + (\alpha/2) T^2$ it is
\begin{equation}
\tilde{\mathcal{T}}_{\mu\nu} = \alpha \, \frac{ g_{\mu\nu} T^2
  - 4 S_{\mu\nu}{}^\lambda \partial_\lambda T} {32\pi(1+\alpha T)} \, .
\label{eq:effset}
\end{equation}
In spherical symmetry the structure of the
above effective stress-energy tensor
is that of an anisotropic perfect fluid of Segr\'{e} characteristic $[1,1,(1,1)]$.
We will therefore refer to its nontrivial
components as the effective energy density,
effective radial pressure, and effective transverse pressure.
In the denominator of (\ref{eq:effset})
one can see that a potential pathology could exist when
$\alpha T = -1$.  However we will confirm below that
pathologies that will arise are not specifically due to this issue.

Extensive investigations were carried out of the properties of the numerical solutions
obtained with various values of $\alpha/M^2$,
with inbound integrations set off from near-infinity
and initial conditions drawn from the Schwarzschild solution. These investigations
revealed three distinct regimes of $\alpha/M^2$
in which the solutions exhibit qualitatively different behavior. We show the different regimes in Fig.~\ref{fig:singsurfrad} and the properties of each regime are summarized below.

\begin{figure}[htbp]
\begin{center}
\includegraphics[width=0.95\columnwidth]{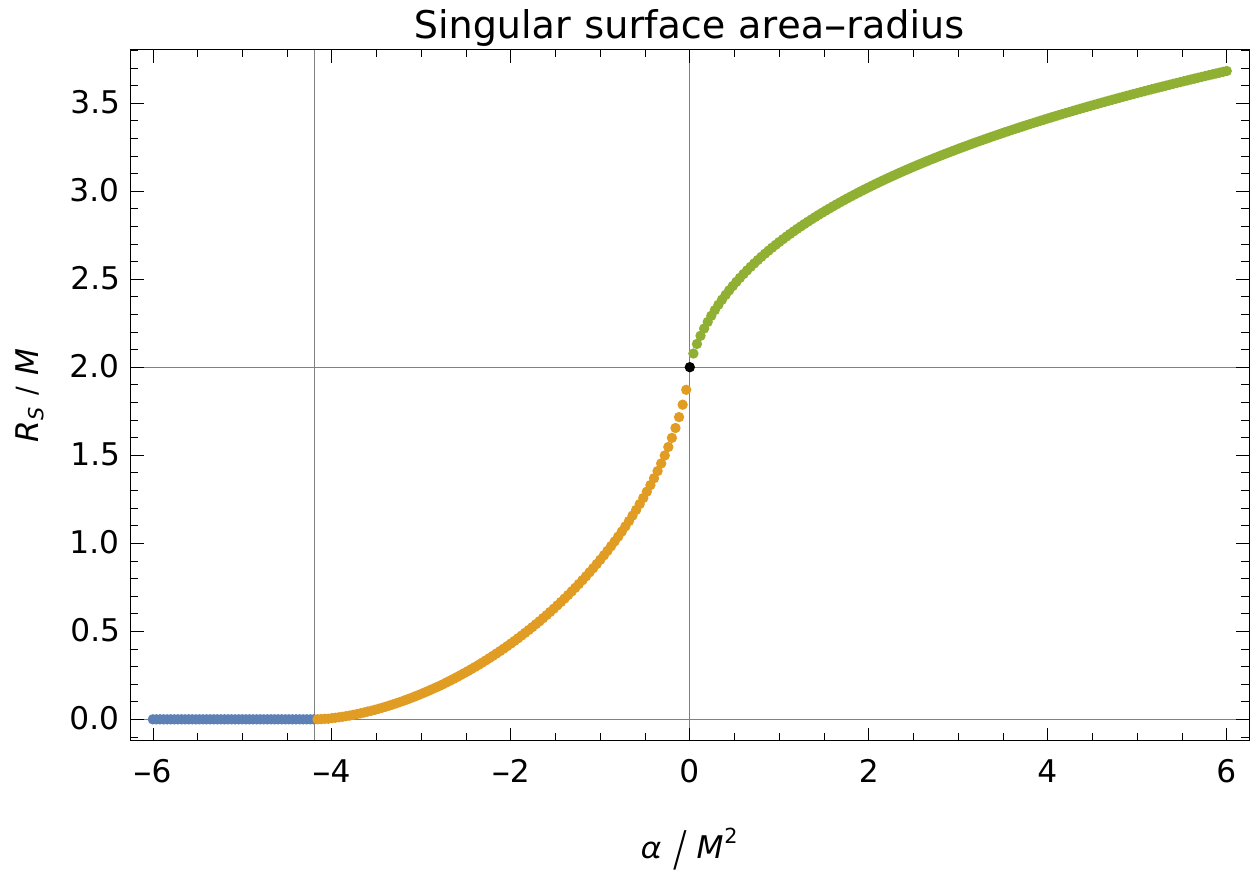}
\end{center}
\vspace*{-5mm}\caption{\label{fig:singsurfrad}{\small Area-radius $R_{\mathrm{S}}$
of singular surfaces in asymptotically Schwarzschild vacuum solutions
in $f(T) = T + \frac \alpha 2 T^2$ gravity:
Point--singularities (blue dots),
infinite gravitational redshift singular surfaces (orange dots),
finite gravitational redshift singular surfaces (green dots).
Black dot at $\alpha=0$ and $R_{\mathrm{S}}=2M$
is the (nonsingular) event horizon of the Schwarzschild black hole.}}
\end{figure}

\begin{itemize}

\item With positive values of $\alpha/M^2$ the numerical routines are able
to carry out the inbound integration of the solutions
only down to a finite value of $x = x_{\mathrm{S}}$
at which a singularity and/or stiffness in the system is reported before a horizon can form.
Functions $A$ and $B$ obtained numerically for $\alpha/M^2=1$
are shown in the upper plot of Fig.~\ref{fig:numfigs}. 
The isotropic radial coordinate of the spherical surface
at which the numerical breakdown occurs is
$\rho_{\mathrm{S}} = M x_{\mathrm{S}} / ( 1 - x_{\mathrm{S}} )$,
and the corresponding area-radius
$ R_{\mathrm{S}} = M x_{\mathrm{S}} / (1 - x_{\mathrm{S}}) $
is shown in Fig.~\ref{fig:singsurfrad} with green dots. The term area-radius refers to the value of the radius that defines the area of 2-spheres. That is, it refers to the corresponding radius in the Schwarzschild coordinates.
We observe that $R_{\mathrm{S}}$ increases with $\alpha/M^2$,
while as $\alpha/M^2 \to 0$ we have $R_{\mathrm{S}} \to 2M$ as expected.
This is shown with green dots in Fig.~\ref{fig:singsurfrad}.

With $\alpha = 0$ one expects the Schwarzschild solution
with the event horizon of radius $R=2M$,
which is an infinite redshift surface.
However, the property of infinite redshift
is not shared with the surfaces $R_{\mathrm{S}}$ obtained with $\alpha/M^2>0$,
as inspection of the solutions prior to the breakdown
reveals that $A(x)$ remains finite
as $x\to x_{\mathrm{S}}$.
Further inspection of the solutions in this regime
reveals that as $x\to x_{\mathrm{S}}$
Ricci--Christoffel and Kretschmann--Christoffel scalars
diverge as can be seen in the upper plot in Fig.~\ref{fig:numdivs}.
This implies that the orthonormal frame Riemann tensor
contains diverging components,
which in general relativity signals diverging tidal forces or infinite geodesic deviations.
Since the motion of particles in $f(T)$ gravity
is governed by the same equations as in general relativity,
we are concluding that the surfaces $R_{\mathrm{S}}$
obtained here involve singular physics.

We also inspected the components
of the effective $f(T)$-fluid stress energy tensor (\ref{eq:effset}).
As $x\to x_{\mathrm{S}}$ it is found to have diverging
energy density and transverse pressure,
while the radial pressure remains finite. These results are also
shown in the upper plot of Fig.~\ref{fig:numdivs}.

\item In the regime $-4.2 \lesssim \alpha/M^2 < 0$,
where the lower boundary of the interval
could only be established approximately,
the numerical solutions can be evolved down to
a finite value of the coordinate $x=x_{\mathrm{S}}$
at which the function $A$ vanishes,
signaling an infinite redshift spherical surface of finite radius
$ R_{\mathrm{S}} = M x_{\mathrm{S}} / (1 - x_{\mathrm{S}}) $
(see orange dots in Fig.~\ref{fig:singsurfrad}).
Metric functions obtained for $\alpha/M^2=-1$ (a representative example of many different values studied in this regime)
are shown in the middle plot of Fig.~\ref{fig:numfigs}. 
Inspection of Ricci--Christoffel and Kretschmann--Christoffel scalars reveals
that they diverge as $x \to x_{\mathrm{S}}$
as illustrated in the middle plot in Fig.~\ref{fig:numdivs}.
As in the previous case, this implies diverging components
of the orthonormal frame Riemann tensor and renders this surface singular
in the sense of infinite tidal forces.

\begin{figure}[!h]
\includegraphics[width=\columnwidth]{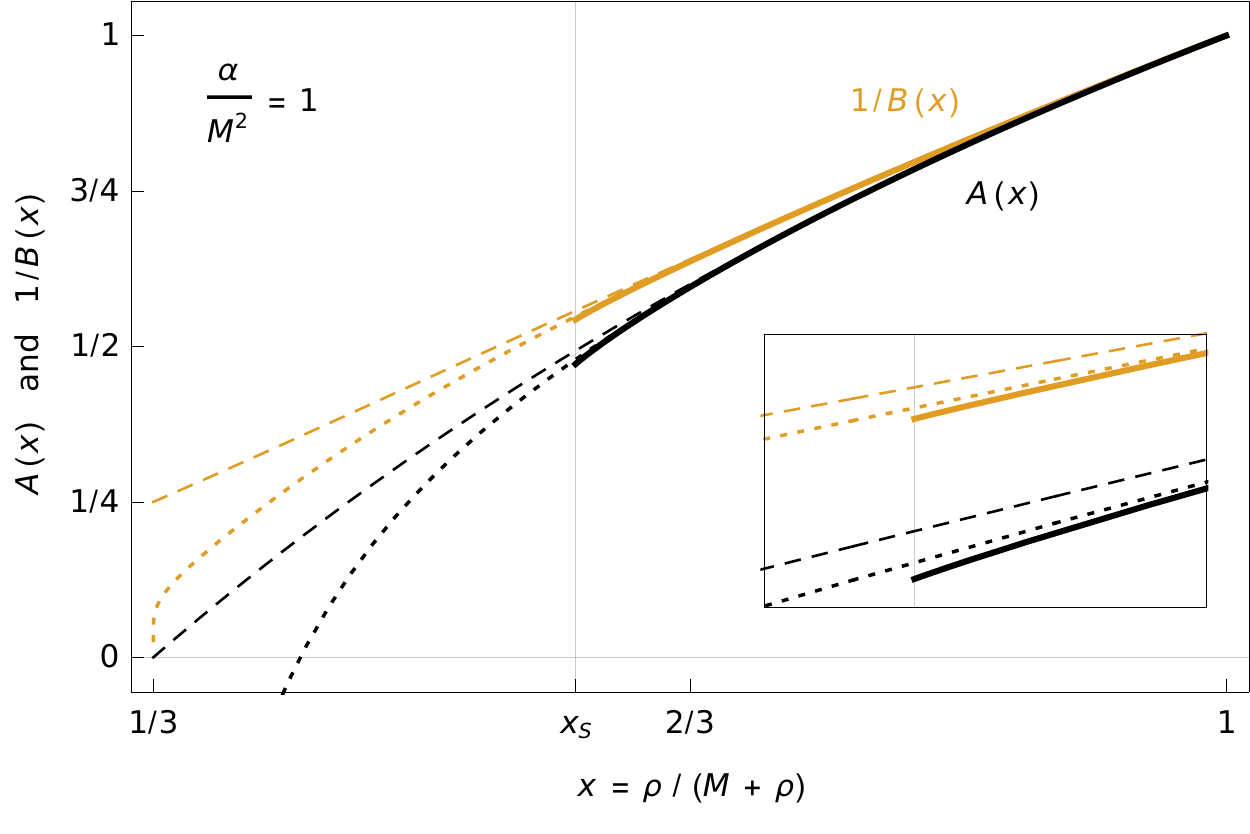} \\[1ex]
\includegraphics[width=\columnwidth]{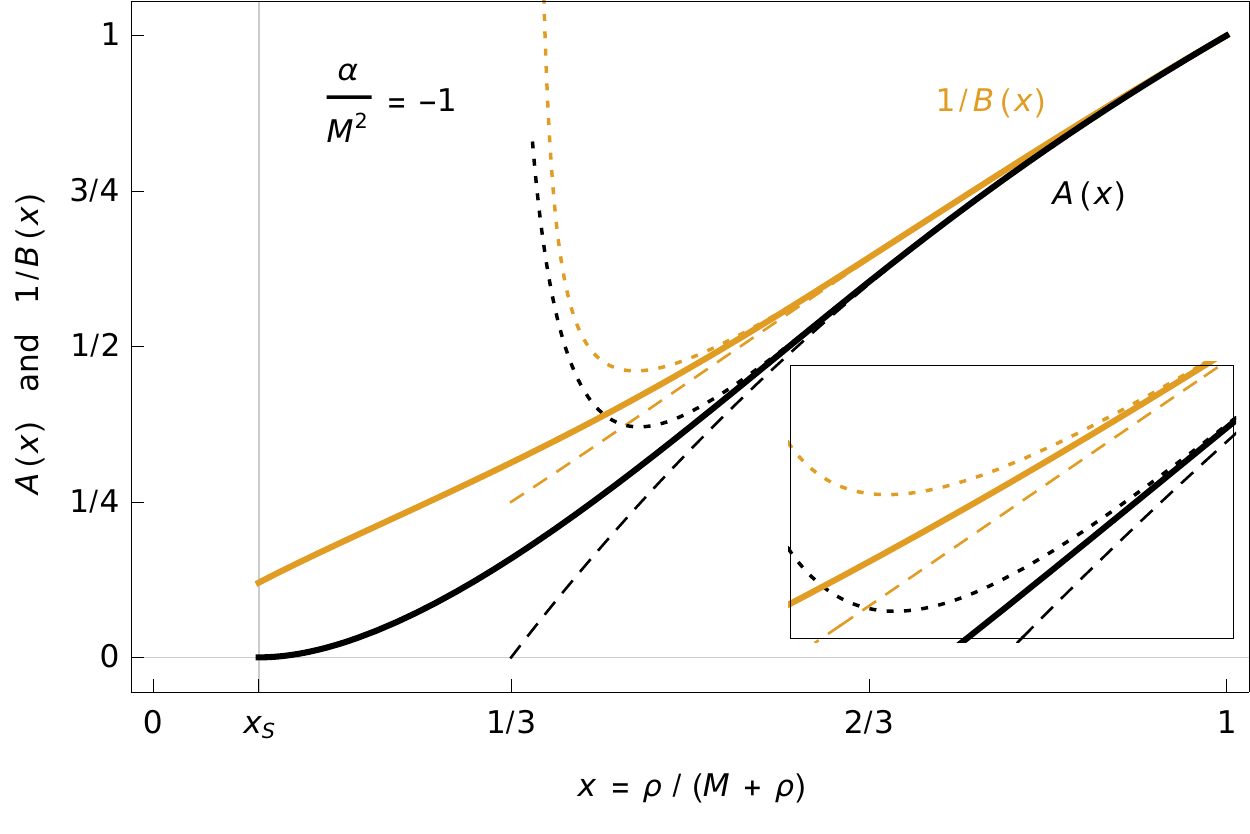} \\[1ex]
\includegraphics[width=\columnwidth]{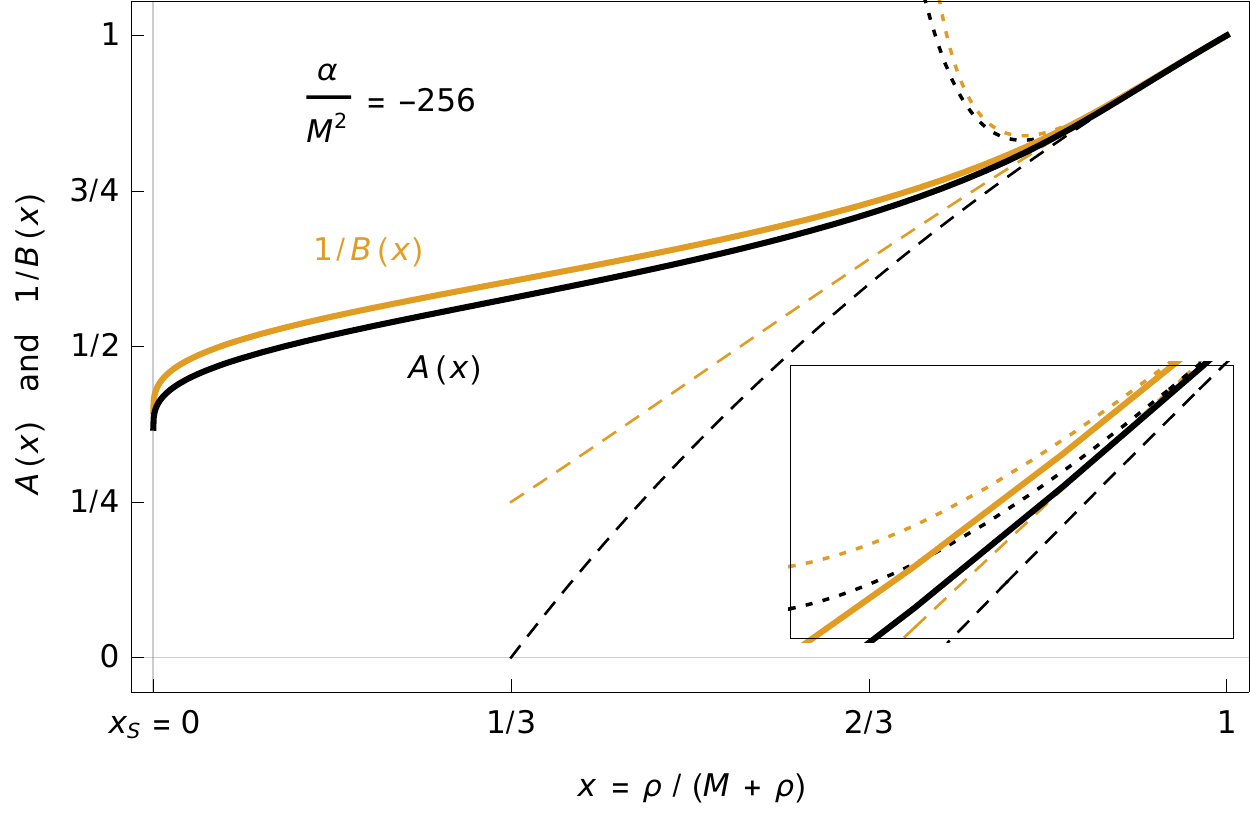}
\vspace*{-8mm}\caption{\label{fig:numfigs} \small
Numerical vacuum solutions in $f(T) = T + \frac \alpha 2 T^2$ gravity:
Functions $A$ (black) and $1/B$ (orange)
obtained numerically (solid lines),
perturbative solutions (short--dashed lines),
Schwarzschild solution (long--dashed lines).
(Insets show that for large $x$ numerical solutions
follow perturbative solution 
more closely than the Schwarzschild solution.)\vspace{-0.3cm}}
\end{figure}

As can be seen from the plots in Fig.~\ref{fig:numfigs},
although $A'(\rho_{\mathrm{S}})=0$,
as was required for nonsingular horizons from our earlier analysis,
we note that the second earlier derived condition required for regularity there,
namely that $A''(\rho_{\mathrm{S}})=0$ as well, does not hold. 

The effective pressures of the $f(T)$-fluid
also diverge as $x \to x_{\mathrm{S}}$,
while the effective energy density remains finite
and this is shown in the middle plot in Fig.~\ref{fig:numdivs}.
Therefore, we consider the surfaces $R_{\mathrm{S}}$
in this regime singular surfaces.

\item In the regime $\alpha/M^2 \lesssim -4.2$
the numerical evolution of solutions can typically be carried out
down to a value of $x \sim 10^{-15}$,
while the exact zero remains out of reach
due to the singularity of coefficients in the differential equations.
These solutions are represented
with blue dots in Fig.~\ref{fig:singsurfrad},
and the metric functions obtained for $\alpha/M^2=-256$
are shown in the lower plot of Fig.~\ref{fig:numfigs}. 
The functions $A$ and $B$ remain finite throughout the range of $x$,
while Ricci--Christoffel and Kretschmann--Christoffel scalars,
as well as the components of the $f(T)$-fluid effective stress energy tensor,
diverge as $x\to 0$.
These are illustrated in the lower plot in Fig.~\ref{fig:numdivs}.
As there is no indication of the divergences taking place
at a finite value of $x$, we  interpret these
solutions as representing a point singularity at the center of symmetry.

\end{itemize}


Based on the results of the numerical evolutions
one can conclude that asymptotically Schwarzschild solutions
obtained with $\alpha\ne0$
contain either a singular central point or a singular surface of finite radius.
These singular surfaces are not shielded by an event horizon
as in the case of the Schwarzschild solution.

It should also be noted that while the $\alpha=0$ (TEGR) case
can in some sense be understood as a legitimate case
within the realm of $f(T) = T + \frac \alpha 2 T^2$ gravity theories,
it is in fact highly special.
This can be seen by observing
that with $\alpha\ne0$ the vacuum equations of motion
reduce to two equations with $A''$ and $B''$ as highest order derivatives,
while with $\alpha=0$ the highest derivatives are $A'$ and $B''$.
Also, in the summary of the numerical results
given in Table \ref{tbl:numvactbl}
one can see that the physical properties of the singular surface
can not be taken as changing continuously as $\alpha/M$ changes sign. One has a singular limit there.

In Table \ref{tbl:numvactbl} we also give the value of the quantity
$(g_{rr})^{-1/2} = 1 + \rho B'(\rho) / B(\rho) $ as $x\to x_{\mathrm{S}}$,
where $g_{rr}$ is the metric component in the usual Schwarzschild chart
whose line element for a purely radial displacement
is $d s^2 = - g_{rr} dr^2$, coordinate $r$ being the area-radius.
In the regime $\alpha/M^2 \lesssim -4.2$
we find that for all solutions $g_{rr} \to 1$ as $x\to 0$,
which is the expected behavior
in spherically symmetric spacetimes with regular centers (e.g.\ stars).
This hints that the divergence
of the effective energy density, $\tilde\rho$, at the center of symmetry
that we find in our numerical solutions
might be sufficiently benign to render the ``mass function'' defined with
$m(r) = r (1-g_{rr}(r)) / 2 = 4\pi \int_0^r r^2 \tilde\rho(r') dr'$
obey the limit $m(r)/r \to 0$ as $r\to 0$,
which is the property of regular centers. However, this is not sufficient to eliminate all pathologies there.

For all the solutions in the regime $-4.2 \lesssim \alpha/M^2 <0$,
as $x\to x_{\mathrm{S}}$ we find $(g_{rr})^{-1/2} \to 1/3$,
for which we have found no direct interpretation, but it is interesting that there exist such attractors in the equations.
For positive values of $\alpha$ the value of $(g_{rr})^{-1/2}$
at the singular surface is found to vary.

We also looked for the possibility that the divergences
in the components of the effective stress-energy tensor
are due to the previously mentioned condition $\alpha T = -1$,
which appears in the denominator of (\ref{eq:effset}).
It was found that this is not the source of the pathology.

\begin{figure}[htbp]
\includegraphics[width=\columnwidth]{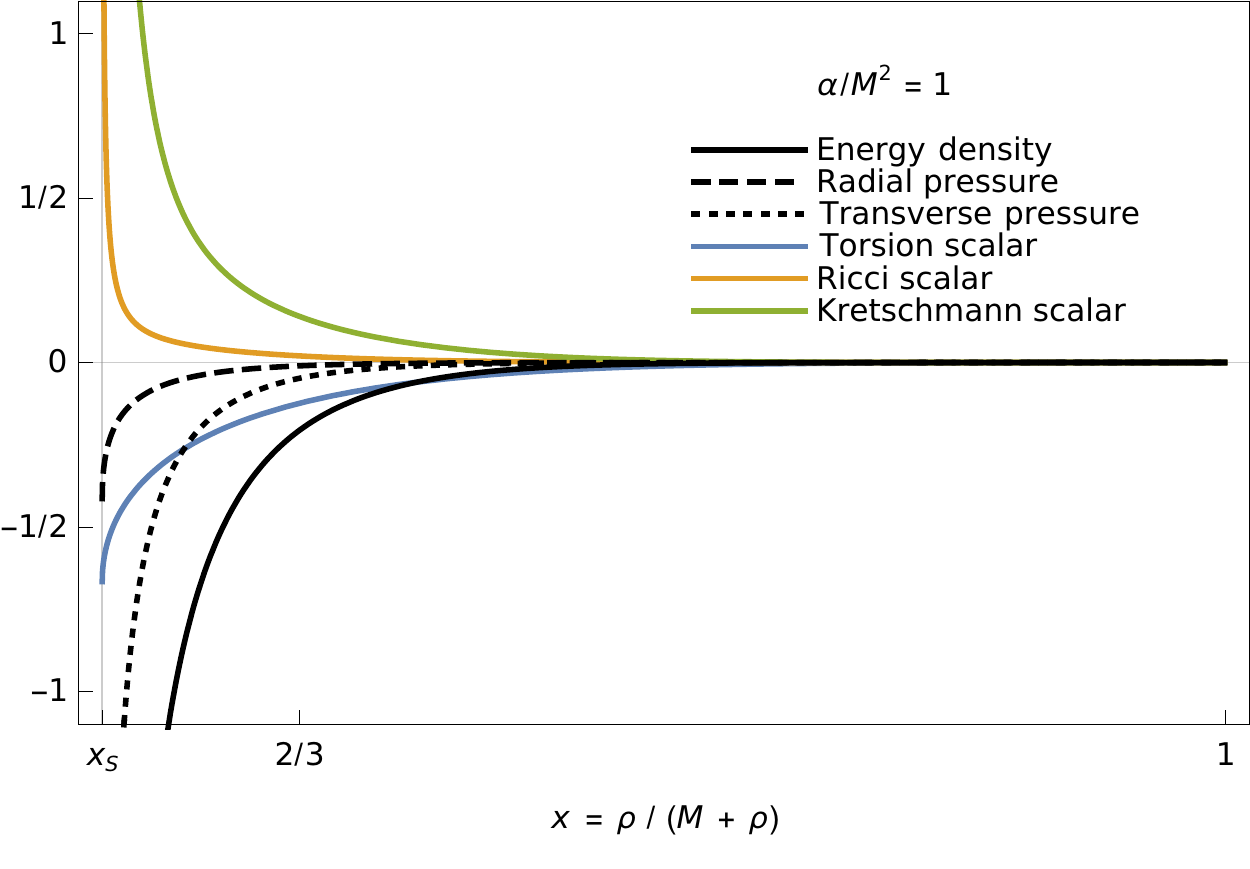} \\[1ex]
\includegraphics[width=\columnwidth]{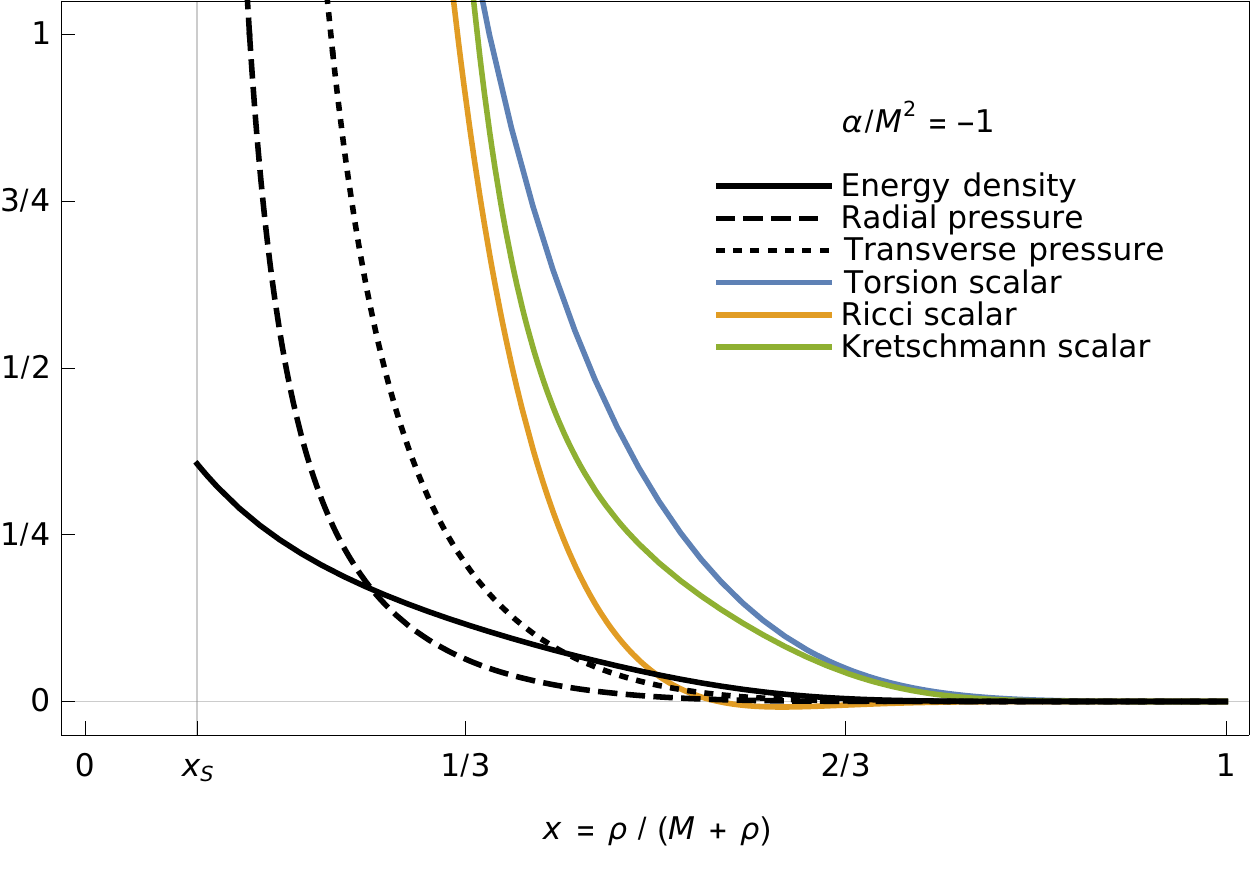} \\[1ex]
\includegraphics[width=\columnwidth]{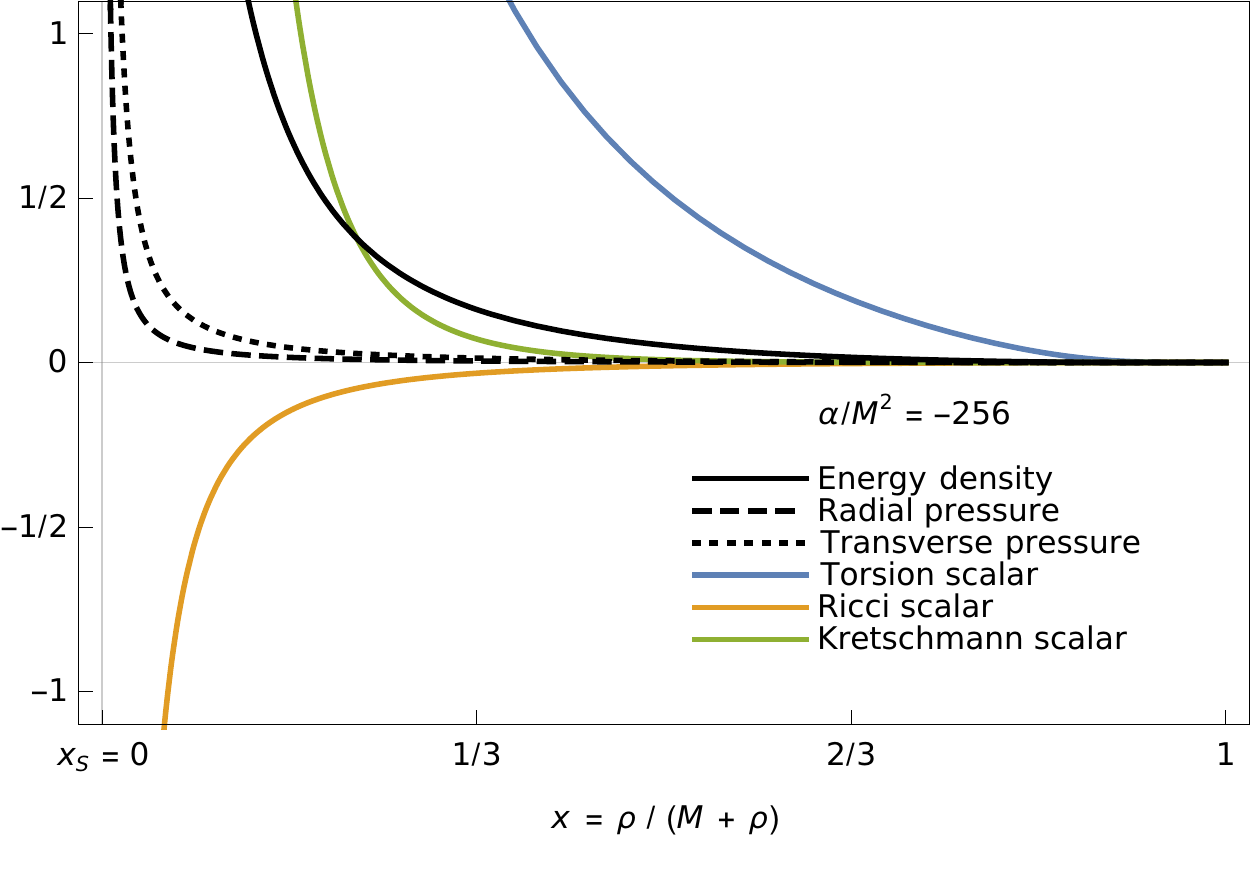} 
\vspace*{-8mm}\caption{\label{fig:numdivs} \small
Effective stress-energy components and invariant scalars
in numerical vacuum solutions in $f(T)$ gravity:
Upper plot: stress--energy components obtained with $\alpha/M^2=1$
are multiplied by $250\,M^2$,
and invariant scalars $\alpha T$, $M^2 \mathring R$,
and $M^4 \mathring R_{\alpha\beta\gamma\delta}
	\mathring R^{\alpha\beta\gamma\delta}$ (Kretschmann scalar),
are multiplied by $2$ for figure scaling purposes.
All quantities except the energy density and the radial pressure
	diverge at the singular surface.
Middle plot: For $\alpha/M^2 = -1$ factor of 10 is used for
stress--energy components, while no additional factors are used
with invariant scalars.  All quantities except the energy density
diverge at the singular surface.
Bottom plot: For $\alpha/M^2 = - 256$ the stress-energy components
are scaled up by factor $10$,
	and invariant scalars are scaled down by factor $10$.
	All quantities diverge at the singular center.}
\end{figure}

\begin{table*}[htbp] 
\caption{\label{tbl:numvactbl}{\setstretch{1.0}
{\small Properties of metric functions $A$ and $B$,
Ricci--Christoffel, torsion, and Kretschmann--Christoffel scalars,
and effective $f(T)$-fluid stress energy tensor components
on singular surfaces (except $\alpha = 0$)
found in numerically constructed vacuum solutions.
Signs of divergences are reported as surface is approached from the outside.
f.p.\ (f.n.) stands for finite positive (negative) value.}}}
\begin{center}
\begin{tabular}{r|c|c|c|c}
Regime & $\alpha/M^2 \lesssim -4.2$
& $-4.2\lesssim\alpha/M^2<0$ & $\alpha = 0$
& $0<\alpha/M^2$ \\ [0.5ex] \hline\hline
Surface condition & num.\ bkdwn & min.\ of $A^2$ & & num.\ bkdwn \\
Isotropic radius $\rho_{\mathrm{S}}$ & $\rho_{\mathrm{S}}\simeq0$ & $0<\rho_{\mathrm{S}}<M/2$ & $M/2$ & $M/2<\rho_{\mathrm{S}}$ \\ \hline
$A(\rho_{\mathrm{S}})$, $A'(\rho_{\mathrm{S}})$, $A''(\rho_{\mathrm{S}})$ & f.p., $+\infty$, $-\infty$ & $\simeq0$, $\simeq0$, f.p. & $0$, $1/M$, $-2/M^2$ & f.p., f.p., f.n. \\
$B(\rho_{\mathrm{S}})$, $B'(\rho_{\mathrm{S}})$ & f.p., $-\infty$ & f.p., f.n. & $4$, $-8/M$ & f.p., f.n. \\
Area--radius $R_{\mathrm{S}} = B(\rho_{\mathrm{S}})\rho_{\mathrm{S}}$ & $0$ & $0<R_{\mathrm{S}}<2M$ & $2M$ & $2M<R_{\mathrm{S}}$ \\
$ 1 + \rho_{\mathrm{S}} B'(\rho_{\mathrm{S}}) / B(\rho_{\mathrm{S}}) $ & $\simeq 1$ & $\simeq 1/3$ & $0$ & f.p. (varies) \\ \hline
Ricci scalar & $ - \infty$ & $ + \infty$ & $0$ & $ + \infty$ \\
Torsion scalar & $ - \infty$ & $ - \infty$ & $-\infty$ & f.n. \\
Kretschmann scalar & $ + \infty$ & $ + \infty$ & $3/4M^2$ & $ + \infty$ \\ \hline
Effective energy density & $ + \infty$ & f.p. & $0$ & $ - \infty$ \\
Radial pressure & $ + \infty$ & $ + \infty$ & $0$ & f.n. \\
Transverse pressure & $ + \infty$ & $ + \infty$ & $0$ & $ - \infty$
\end{tabular}
\end{center}
\end{table*}

\section[Some comments on the cosmological constant]{Some comments on the cosmological constant} \label{sec:cosconst}

We make some comments here regarding the possibility
of a non-zero cosmological constant, $\Lambda$.
This can be accommodated by adding
a $-\Lambda h_{\hat{\alpha}\mu} h^{\hat{\alpha}}{}_{\nu}$ term
to the left-hand side of (\ref{eq:eoms}).
This term of course could be interpreted as the ``stress-energy of the vacuum''
if one moves it to the right-hand-side of the equations; that is
$\mathcal{T}^{\mu}{}_{\nu\,\mbox{\tiny{(vac)}}}
:= \Lambda/(8\pi)\delta^{\mu}{}_{\nu}$.

Regarding the regularity at the center, it is still demanded that $A^{\prime}(0)$ and $B^{\prime}(0)$ vanish as previously required, since those conditions arose not from the equations of motion, but from a reasonable force equation at the center (\ref{eq:geoeqn}).  If one performs the Taylor expansions done previously about $\rho=0$, but now with the equations supplemented with a cosmological term, solving order-by-order reveals that the restriction $B^{\prime\prime}(0)=0$ no longer holds. Instead, $B^{\prime\prime}(0)$ now must be set to a function of $\Lambda$ and $B(0)$. This now implies that even under the restriction of regularity everywhere, the spacetime is no longer Minkowski. This is not surprising of course given the non-zero cosmological term.

With regards to the issue of horizons, the argument that any solution to the equations of motion near a horizon require that $A^{\prime}(\rhoh)B^{\prime}(\rhoh)=0$ still holds with a cosmological term. Therefore the analysis on the orthonormal Riemann-Christoffel components (\ref{eq:riemangt}) and (\ref{eq:riemrt}) still holds as before. That is $A^{\prime}(\rhoh)$ and $A^{\prime\prime}(\rhoh)$ should equal zero in order for the horizon to be non-singular. However, in the case of a cosmological constant we do not have numerical confirmation that $A^{\prime\prime}(\rhoh)$ does not equal zero on possible horizons.
This impediment arises from the fact
that the numerical solution requires an asymptotic
weak-field solution in order to commence the computational evolution.
Asymptotically one could use the Schwarzschild-(anti)de Sitter solution
for this, however, this is problematic
since the cosmological horizon
renders the radial coordinate timelike far from the center
\cite{paperonschdesitterinisotropiccoordiates}.
%

If we are willing to restrict the horizon analysis to analytic functions,
we can perform a Taylor analysis about $\rho=\rhoh$
on the equations of motion subject to the found conditions
that $A^{\prime}(\rhoh)$ and $A^{\prime\prime}(\rhoh)$ must be zero
for regularity there.
These conditions are not reliant on any analyticity restriction on the tetrad so apply in general to any horizon. The pattern that arose for the previous no-go result is no longer found in the case where $\Lambda$ is present in the equations of motion. Instead we were able to solve the equations up to some order $(\rho-\rhoh)^{n}$. For $L_{\hat{t}\hat{t}}$ we solved up to and including order $n=4$, for $L_{\hat{\rho}\hat{\rho}}$ to order $n=3$, and for $L_{\hat{\theta}\hat{\theta}}$ to $n=3$. A consistent solution could be found up to these orders subject to the following extra conditions:
\begin{equation}
B^{\prime}(\rhoh) = 0 \,, \qquad
\Lambda = - \frac{1}{4\alpha}\,. \label{eq:lambcond2}
\end{equation}
The restriction on $B^{\prime}(\rhoh)$ in (\ref{eq:lambcond2}) is not present in TEGR. The reason is that the leading order terms in the equations of motion are proportional to $\alpha B^{\prime}$ whose only consistent solution is that this must vanish. For $\alpha=0$, of course, this condition is automatically met, but for non-zero alpha we must enforce $B^{\prime}(\rhoh)=0$. We hasten to add here that we were not able to show
that a non-singular horizon can exist
with the presence of non-zero $\Lambda$,
but simply that a non-existence argument could not be easily formulated in this case, and \emph{if} they exist conditions (\ref{eq:lambcond2}) must hold, at least assuming the system is describable by analytic functions.

The second condition in (\ref{eq:lambcond2})
is somewhat interesting in that it implies
that if regular horizons exist the sign of the cosmological constant
is tied to the sign of the nonlinear torsion coupling, at least if subject to the condition of analyticity.


\section{Concluding remarks}

In this manuscript properties of the spherically symmetric static vacuum in the minimal quadratic extension to TEGR in $f(T)$ gravity theory were studied. A number of interesting results were found. It is determined that demanding vacuum regularity at the center, the center being the point of isotropy, requires that the first derivatives of the the tetrad function vanish. The field equations under the symmetry then dictate that the spacetime is Minkowski spacetime throughout. Although it may seem that this must be the case, it is not completely a priori obvious, since $f(T)$ gravity mimics general relativity in the presence of an exotic material source, as can be seen from (\ref{eq:eoms}), and such a result of global Minkowski spacetime is of course not required in non-vacuum general relativity. 

The properties of possible vacuum horizons were also studied. The situation here differs from the above central analysis in that one no longer makes the demand that the spacetime be regular everywhere in order to be physically acceptable, but instead only regular up to and including the horizon. It was found that, with the exception of TEGR, vacuum horizons possess some type of pathology in the theory.
Namely, for asymptotically Schwarzschild solutions with $\alpha>0$
horizons cannot exist, and for $\alpha < 0$ there is a range of $\alpha$
where they may exist, but are singular.
We therefore have naked singularities evading the cosmic censorship conjecture.
These results do not rely on the non-linear torsion term being small,
and so are fairly general,
although we do confirm the validity of the perturbative solution
derived in \cite{ref:ourvac} in the weak-field regime of the exact theory.

We should add here though that the results, although quite interesting,
do not mean that black holes are completely forbidden
within the studied theory.
It is possible that relaxing the symmetry to stationary
instead of static will re-introduce physical (non-singular) horizons.
Or else time dependence, even within spherical symmetry, may allow for non-singular horizons, recalling that as of yet there is no Birkhoff's theorem forbidding this for extended teleparallel gravity theory. At the moment these extensions pose a difficult task as there is no direct way to calculate the appropriate inertial spin connection for these scenarios. 
Also, some of the obtained no--go results are based
on the assumption that the solution is asymptotically Schwarzschild.
Abandoning this assumption could potentially lead
to nonsingular horizons in some cases where the horizon cannot be described by analytic functions.
However, such solutions, if they exist,
would be astrophysically less interesting. There is also the possibility of adding matter to the system, which could allow evasion of the horizon no-go results here which apply to the vacuum case only. It is also possible that the correct theory of gravity is not the extended $f(T)=T+\alpha T^{2}/2$ gravity but instead generalizations on it such as $f(T,B)$ \cite{ref:bbh} or other extended teleparallel theories. These extensions were not studied here.

The analytic work in the first sections of this manuscript was supplemented with computational work in section \ref{sec:computational} and the computational work confirms all of the obtained results. In no scenario, save for TEGR ($\alpha=0$), were regular horizons found. 

Finally, some comments were made regarding the possible addition of a cosmological constant. It was found that a non-existence argument is more difficult to formulate and the analysis of the possible existence of a regular vacuum horizon was inconclusive. It might therefore be possible that regular vacuum static spherically symmetric horizons exist with non-zero $\Lambda$. 

\vspace{0.7cm}
\section*{Acknowledgments}
This work is partially supported
by the VIF program of the University of Zagreb.
We thank D.~Horvat and Z.~Naran{\v c}i{\'c}
for fruitful conversations related to this work. We also thank the anonymous referee for comments that have helped clarify the manuscript.

\PRLsep
\vspace{-0.080cm}

\linespread{0.6}
\bibliographystyle{unsrt}
 

%



\end{document}